\documentclass[11pt]{article}

\usepackage[final]{acl}

\usepackage{times}
\usepackage{latexsym}

\usepackage[T1]{fontenc}

\usepackage[utf8]{inputenc}

\usepackage{microtype}

\usepackage{inconsolata}

\usepackage{graphicx}

\usepackage{amsmath}

\usepackage{amsthm}

\usepackage{algorithm}
\usepackage{algorithmic}

\usepackage{tcolorbox}
\usepackage{enumitem} 

\usepackage{booktabs}
\usepackage{multirow}
\usepackage{xcolor}
\usepackage{array}
\usepackage[table]{xcolor}

\usepackage{pifont}

\definecolor{yifan}{RGB}{30, 80, 190}
\newcommand{\yifan}[1]{\textcolor{black}{#1}}
\newcommand{\yaokun}[1]{\textcolor{black}{#1}}
\usepackage[dvipsnames]{xcolor}

\usepackage{dashrule}

\newtheorem{assumption}{Assumption}

\newtheoremstyle{propositionstyle}
  {3pt}   
  {3pt}   
  {\normalfont} 
  {}      
  {\bfseries} 
  {.}     
  {0.5em} 
  {}      

\theoremstyle{propositionstyle}
\newtheorem{proposition}{Proposition}

\title{PropUQ-MAS: Propagation-Aware Uncertainty Quantification for LLM Multi-Agent Systems}

\author{
\textbf{Yaokun Liu\textsuperscript{1*}} \quad
\textbf{Yifan Liu\textsuperscript{1*}} \quad
\textbf{Daniel Yue Zhang\textsuperscript{2}} \quad
\textbf{Ruichen Yao\textsuperscript{1}} \\
\textbf{Zelin Li\textsuperscript{1}} \quad
\textbf{Dong Wang\textsuperscript{1}}
\\
\textsuperscript{1}University of Illinois Urbana-Champaign
\\
\textsuperscript{2}Scale AI
\\
\texttt{\{yaokunl2, yifan40, ryao8, zelin3, dwang24\}@illinois.edu}
\\
\texttt{yue.zhang@scale.com}
}

\begin{document}
\maketitle

\begingroup
\renewcommand{\thefootnote}{*}
\footnotetext{Equal contribution.}
\renewcommand{\thefootnote}{1}
\footnotetext{Code is released at \url{https://github.com/yaokunliu/PropUQ-MAS.git}.}
\endgroup

\begin{abstract}

LLM-based multi-agent systems (MAS) solve complex tasks through communication among role-specialized agents. However, inter-agent dependencies introduce reliability risks beyond isolated agent failures. For instance, errors in intermediate messages could be inherited and amplified by downstream agents. 
Existing uncertainty quantification (UQ) methods mainly target isolated responses or single-agent reasoning, and therefore fail to capture uncertainty propagation in MAS. To this end, we propose \textsc{PropUQ-MAS}, an error propagation-aware UQ framework that represents MAS execution as a communication-structured graph and estimates each step's reliability by combining local uncertainty with uncertainty inherited from upstream messages. 
Extensive experiments demonstrate that \textsc{PropUQ-MAS} consistently improves UQ in MAS, with average relative gains of $+6.10\%$ in AUROC and $+47.58\%$ in PRR.\textsuperscript{1}

\end{abstract}

\section{Introduction}

Large language model (LLM)-based multi-agent systems (MAS) have emerged as a powerful paradigm for solving complex tasks by orchestrating role-specialized agents across multi-step workflows~\cite{wu2024autogen,li2023camel,hong2024metagpt}. Through inter-agent communication and strategic collaboration, MAS exhibit superior flexibility in demanding domains such as autonomous software engineering and multi-step web navigation~\cite{liu2024agentbench}.
\yifan{However, the same inter-agent dependencies that enable such flexibility can also shift the reliability bottleneck from individual agent failures to system-wide error propagation.}

Unlike isolated agents whose errors stem primarily from intrinsic limitations (e.g., hallucinations), MAS reliability is a collective property governed by interaction dynamics: \yaokun{An erroneous intermediate message from an upstream agent may be interpreted as a valid context by downstream agents, incorporated into subsequent reasoning, and propagated or even amplified through inter-agent communication~\cite{position1,position2}.}

\begin{figure}[t]
    \centering
    \includegraphics[width=1\columnwidth]{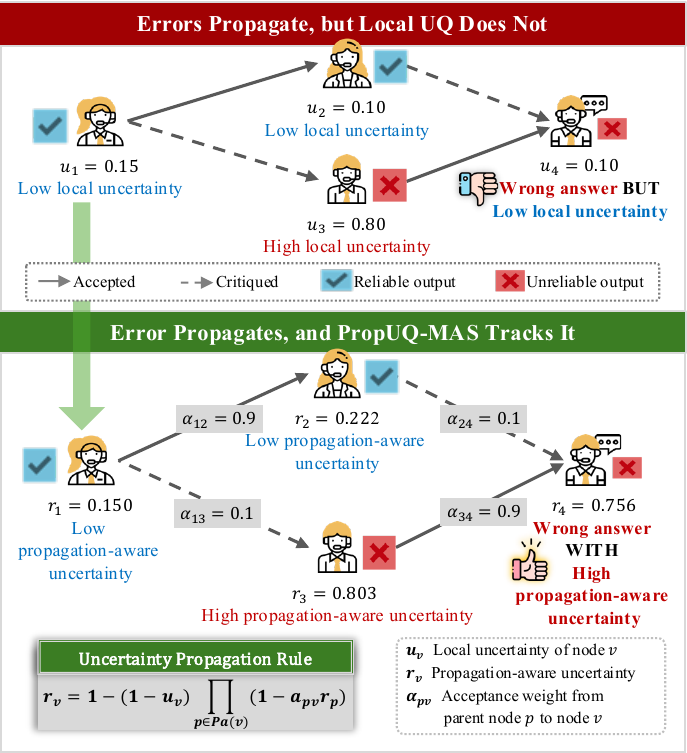}
    \caption{
    Local UQ misses propagated risk, while \textsc{PropUQ-MAS} tracks propagation-aware uncertainty over the MAS execution graph.
    }
    \label{fig:intro}
\end{figure}

To mitigate the cascading risks, we argue that MAS requires a step-wise propagation-aware uncertainty quantification (UQ) framework,
\yaokun{which estimates the marginal failure risks of each intermediate output during the MAS execution trajectory.}
\yifan{Such a framework can serve as a diagnostic signal for MAS by supporting real-time monitoring of risky intermediate states, adaptive intervention before errors accumulate, and post-hoc attribution of failure-contributing steps~\cite{shinn2023reflexion,zhang2025which}. 
These target applications require the UQ framework to be topology-agnostic and online-computable: it should handle diverse MAS structures and update uncertainty as each intermediate output is generated, without waiting for the full trajectory to complete.}

\yifan{However, existing UQ frameworks are not designed to track step-wise uncertainty propagation in MAS. As illustrated in Figure~\ref{fig:intro}, traditional LLM UQ methods estimate local uncertainty for isolated single-turn outputs using signals such as token probabilities~\cite{nll,ccp}, self-evaluation~\cite{ask4conf1,ptrue}, semantic consistency~\cite{selfcheckgpt,se,liu2025reasoning}, or internal hidden representations~\cite{liu2026mind}. 
Recent agentic UQ methods extend local uncertainty estimation from single responses to multi-step reasoning trajectories~\cite{zhao2025uncertainty,duan2025uprop}. These methods assume a sequential single-agent trajectory, whereas MAS introduces more diverse communication topologies.
These differences introduce MAS-specific UQ challenges that are not addressed by agentic UQ, yet MAS-specific UQ remains largely under-explored. More recently, MATU~\cite{chen2026every} estimates trajectory-level uncertainty by the consistency of multiple sampled executions. This post-mortem design neither models uncertainty propagation nor supports real-time node-wise UQ for online monitoring.}

To this end, we propose \textsc{PropUQ-MAS}, a propagation-aware MAS UQ framework for step-wise reliability estimation. 
By unfolding MAS trajectories into a directed acyclic graph, \textsc{PropUQ-MAS} bypasses structural constraints to deliver a topology-agnostic framework. 
Building on this graph view, we derive a recursive propagation rule with a probabilistic interpretation: each node's uncertainty is updated by combining its local uncertainty with propagated uncertainty inherited from upstream agents. 
The inherited component explicitly modulates the uncertainty propagation strength over communication channels, formalizing the likelihood of error contamination across different agent behaviors, including acceptance and critique.
This establishes a tractable propagation rule that quantifies uncertainty seamlessly in a single forward pass. 
Extensive experiments demonstrate that \textsc{PropUQ-MAS} consistently improves step-wise UQ in MAS, with average relative gains of $+6.10\%$ in AUROC and $+47.58\%$ in PRR, while showing strong generalization and interpretability.

\section{Related Work}
\subsection{LLM Uncertainty Quantification}

Existing LLM UQ methods estimate the reliability of an individual generation using token-level likelihoods~\cite{nll,ccp,lm-polygraph}, verbalized confidence or self-evaluation~\cite{ptrue,ask4conf1,ask4conf2}, sampling-based semantic consistency~\cite{se,selfcheckgpt}, and internal hidden representations~\cite{liu2026mind}. These methods provide useful local reliability signals, which we use as the local uncertainty of each node. However, they treat each generation as an isolated prediction target and do not model how errors can be inherited through later interactions. As a result, a downstream output may appear locally confident while still being unreliable due to a contaminated upstream context.

Recent agentic UQ methods further extend local uncertainty estimation to multi-step single-agent trajectories, either by learning situation-aware step weights or decomposing sequential decision uncertainty~\cite{zhao2025uncertainty,duan2025uprop}. 
Although these methods model uncertainty over sequential trajectories, they cannot directly transfer to complex MAS topologies, such as hierarchical structures with multi-parent or multi-recipient communication. In contrast, \textsc{PropUQ-MAS} models uncertainty propagation over execution graphs and applies to arbitrary interaction topologies.

\subsection{MAS Uncertainty Quantification}

Recent work shows that MAS reliability depends on communication dynamics~\cite{position1,position2,zhang2025which,position3}, motivating step-wise reliability monitoring. However, MAS-level UQ remains largely under-explored, and existing studies do not explicitly model how uncertainty propagates across agents during execution. MATU estimates trajectory-level uncertainty by measuring consistency across multiple sampled MAS executions~\cite{chen2026every}. While useful for holistic reliability assessment, such a post-hoc design requires repeated executions and does not explicitly model how uncertainty is transmitted through inter-agent communication, making it less suitable for real-time node-wise monitoring.
In contrast, \textsc{PropUQ-MAS} is a real-time, training-free propagation layer that computes node-wise UQ in a single MAS forward pass, enabling online monitoring of intermediate outputs.


\section{Methodology}

\subsection{\yifan{Problem Setup}}
\label{sec:problem_formulation}

The execution trajectory of a multi-agent system with arbitrary topology
can be represented as a directed acyclic execution graph (DAG) $\mathcal{G}=(\mathcal{V},\mathcal{E})$.
Each node $v=(i,t)\in\mathcal{V}$ denotes an output (e.g., an intermediate message, tool-use result, or final response) produced by agent $i$ at step $t$. Each directed edge $(p,v)\in \mathcal{E}$ indicates that the generation of node $v$ conditions on the information from node $p$. The parent set of $v$ is denoted as $\mathrm{Pa}(v)=\{p:(p,v)\in\mathcal{E}\}$. 
\yifan{
For MAS with feedback or repeated communication, we unroll the observed execution in temporal order: each new agent output becomes a separate node, and edges point from earlier outputs used as context to the later output they condition. Thus, even recurrent communication patterns form a DAG over the realized execution trace.
Unlike a single-agent trajectory, which usually follows one temporal chain, an MAS execution graph allows a node to have multiple parents when an agent conditions on several upstream messages.}


In this paper, our goal is to quantify propagation-aware uncertainty as a reliability signal for each intermediate output in the MAS execution graph. For each node $v$, let $E_v\in\{0,1\}$ denote its error event, where $E_v=1$ indicates the output is incorrect. The propagation-aware uncertainty of node $v$ is then defined as the marginal error probability:
\[
r_v := \Pr(E_v=1), \quad \forall v \in \mathcal{V}.
\]




\subsection{Uncertainty Propagation in MAS}

\yifan{In MAS execution, node uncertainty does not arise only from local agent generation, but can also be inherited from upstream agents through inter-agent communication. 
We therefore focus on the propagated component of node-wise uncertainty, which depends on the accumulated uncertainties of parent nodes as a recursive mapping $\mathcal{F}$:
\[
r_v = \mathcal{F} \left( \{ r_p\}_{p \in \mathrm{Pa}(v)} \right),
\]
where $r_p$ represents the accumulated uncertainty from parent node $p$.}

\subsubsection{\yifan{Error Event Decomposition}}
\label{sec:event_definitions}

\yifan{To instantiate the recursive mapping $\mathcal{F}$, we introduce an event-level decomposition over the execution graph $\mathcal{G}$, separating errors caused by local agent generation from errors transmitted through incoming interactions. }
For each node $v$, we decouple its uncertainty into a local component and a cascading component inherited from parent nodes through inter-agent interactions.

\paragraph{Local Errors.} The local error event is denoted by $I_v\in\{0,1\}$, where $I_v=1$ represents that node $v$ produces an incorrect output by the inherent reasoning or generation flaws of agent $i$ at step $t$, isolated from the correctness of its parent inputs. We define the local uncertainty of node $v$ as:
\[
u_v := \Pr\left(I_v=1\right),
\]
where $u_v\in[0,1]$ can be estimated by off-the-shelf single-agent uncertainty quantification methods
\yifan{, including verbalized self-evaluation~\cite{ask4conf1,ptrue}, token-level predictive probabilities~\cite{nll,ccp}, and sampling-based self-consistency~\cite{selfcheckgpt,se}.}

\paragraph{Propagated Errors.} \yifan{To formalize error propagation in an MAS, we treat each edge $(p,v)$ as a potential channel of error transmission.} For each edge $(p,v)\in\mathcal{E}$, node $v$ processes information from parent node $p$ in one of two mutually exclusive modes: acceptance ($\mathrm{A}$) or critique ($\mathrm{C}$). Let $M_{pv}\in\{\mathrm{A},\mathrm{C}\}$ denote this edge-level mode. Conditioned on parent error $E_p=1$, we define:
\begin{equation*}
\begin{gathered}
\alpha_{pv} := \Pr(M_{pv} = \mathrm{A} \mid E_p = 1),\\
\beta_{pv}:=\Pr(M_{pv}=\mathrm{C} \mid E_p = 1), \\
\quad \text{s.t. } \alpha_{pv} + \beta_{pv} = 1,
\end{gathered}
\end{equation*}
where $\alpha_{pv},\beta_{pv}\in[0,1]$ denote the probabilities that node $v$ accepts or critiques an erroneous output from parent node $p$, respectively. 
\yaokun{In online execution, the true error-conditioned parameter $\alpha_{pv}$ is not directly observable as $E_p$ is unknown. We therefore instantiate $\alpha_{pv}$ with a self-reported acceptance score $\hat{\alpha}_{pv}$, which serves as a plug-in estimate of the edge-level transmission strength. More implementation details are provided in Appendix~\ref{app:self_reporting}.}


\yifan{Based on these two interaction modes}, we define two edge-level events that characterize how errors are transmitted or blocked along an interaction edge.
The contamination event $Z_{pv}$ occurs when the parent node $p$ is erroneous and $v$ accepts its information. The correction event $Y_{pv}$ occurs when parent node $p$ is erroneous and $v$ critique it:
\begin{equation*}
\begin{gathered}
Z_{pv} := \mathbf{1}\{E_p=1,\,M_{pv}=\mathrm{A}\}, \\
Y_{pv} := \mathbf{1}\{E_p=1,\,M_{pv}=\mathrm{C}\}.
\end{gathered}
\end{equation*}
Given $r_p=\Pr(E_p=1)$, the corresponding contamination and correction probabilities are:
\begin{equation*}
\Pr(Z_{pv}=1) = \alpha_{pv} r_p, \quad \Pr(Y_{pv}=1) = \beta_{pv} r_p.
\end{equation*}

Structurally, the overall error event $E_v$ at node $v$ occurs if and only if the agent either suffers from a local reasoning failure or is contaminated by at least one erroneous parent:
\begin{equation}
\label{eq:error_logic}
E_v = I_v \vee \left( \bigvee_{p \in \mathrm{Pa}(v)} Z_{pv} \right).
\end{equation}
The correction event $Y_{pv}$ does not appear in $E_v$ because it is not an error-generating event.

\subsubsection{\yifan{Recursive Uncertainty Propagation}}
\label{sec:assumptions}

The event decomposition in Eq.~\eqref{eq:error_logic} directly implies that node $v$ is correct if and only if it does not suffer from a local error and none of its incoming edges transmit contamination:
\[
\Pr(E_v=0)
=
\Pr\left(I_v=0,\bigwedge_{p\in\mathrm{Pa}(v)} Z_{pv}=0\right).
\]
However, computing this joint probability exactly would require the joint distribution over local errors and incoming contamination events, which is not available during online MAS execution. We therefore introduce a propagation layer that only requires marginal parent uncertainties and edge-level acceptance probabilities, leading to the following conditional independence assumptions.

\begin{assumption}[Edge-wise conditional independence]
\label{asmp:edge_independence}
At node $v$, the incoming contamination events are represented by their marginal probabilities and treated as independent:
\[
\Pr\left(\bigwedge_{p\in\mathrm{Pa}(v)} Z_{pv}=0\right)
=
\prod_{p\in\mathrm{Pa}(v)} \Pr(Z_{pv}=0).
\]
\end{assumption}

\begin{assumption}[Local-propagation independence]
\label{asmp:intrinsic_independence}
The local error event $I_v$ is independent of the incoming contamination events $\{Z_{pv}\}_{p\in\mathrm{Pa}(v)}$ for each node $v$.
\end{assumption}

Under Assumptions~\ref{asmp:edge_independence} and~\ref{asmp:intrinsic_independence}, the joint probability factorizes as:
\[
\Pr(E_v=0)
=
\Pr(I_v=0)
\prod_{p\in\mathrm{Pa}(v)}\Pr(Z_{pv}=0).
\]
Therefore, we obtain the node-wise uncertainty recurrence:
\begin{equation}
\label{eq:local_recurrence}
r_v
=
1-(1-u_v)\prod_{p\in\mathrm{Pa}(v)}(1-\alpha_{pv}r_p).
\end{equation}
A formal derivation of Eq.~\eqref{eq:local_recurrence} is provided in Appendix~\ref{app:local_recurrence_derivation}. Eq.~\eqref{eq:local_recurrence} gives the node-wise uncertainty propagation rule for MAS execution graphs. This recurrence updates the uncertainty of each agent at each step by combining its local uncertainty with the propagated risks.


\begin{algorithm}[t]
\caption{\yifan{Propagation-aware MAS UQ}}
\label{alg:online_propagation}
\begin{algorithmic}[1]
\REQUIRE Execution graph $\mathcal{G}=(\mathcal{V},\mathcal{E})$.
\ENSURE Propagated uncertainties $\{r_v\}_{v\in\mathcal{V}}$.
\STATE Obtain a topological ordering of $\mathcal{G}$.
\FOR{each node $v$ in topological order}
    \STATE Estimate local uncertainty $u_v$ using a single-agent UQ method.
    \FOR{each parent $p\in\mathrm{Pa}(v)$}
        \STATE Estimate $\hat{\alpha}_{pv}$ by model self-reporting.
    \ENDFOR
    \IF{$\mathrm{Pa}(v)=\emptyset$}
        \STATE $r_v \leftarrow u_v$.
    \ELSE
        \STATE $r_v \leftarrow 1-(1-u_v)\prod_{p\in\mathrm{Pa}(v)}(1-\hat{\alpha}_{pv}r_p)$.
    \ENDIF
\ENDFOR
\RETURN $\{r_v\}_{v\in\mathcal{V}}$.
\end{algorithmic}
\end{algorithm}

\subsection{\yifan{Online Inference and Key Properties}}
\label{sec:properties}

\yifan{As illustrated in Algorithm~\ref{alg:online_propagation}, \textsc{PropUQ-MAS} instantiates the recursive propagation rule as an online reliability tracking procedure over the MAS execution graph.} Since arbitrary MAS interaction structures can be unfolded into a DAG-structured execution trace, the same procedure applies across different MAS topologies. 
Given $\{u_v\}_{v\in\mathcal{V}}$ and $\{\hat{\alpha}_{pv}\}_{(p,v)\in\mathcal{E}}$, node-wise uncertainties are evaluated on-the-fly using Eq.~\eqref{eq:local_recurrence}.
Crucially, since the uncertainty recurrence depends solely on immediate predecessors, $r_v$ can be computed in a streaming fashion during MAS execution without requiring full trajectory completion. These propagated uncertainties serve as real-time reliability signals directly attached to each agent's intermediate messages at each step, which empower proactive downstream control policies, such as dynamic localized replanning, routing low-confidence steps to external verifiers, or executing early stopping to terminate unpromising paths for reliable MAS.

\paragraph{Properties of \textsc{PropUQ-MAS}.}\yifan{Beyond online computability and topology-agnostic applicability, Eq.~\eqref{eq:local_recurrence} also satisfies several desirable provable properties demonstrating its probabilistic validity, non-accumulative under critique, and computational efficiency, which support its use as a propagation-aware UQ layer.}
Detailed proofs are provided in Appendix~\ref{app:properties}.

\begin{proposition}[\textbf{Boundedness}]
\label{prop:boundedness}
For any node $v \in \mathcal{V}$, if $u_v, r_p, \alpha_{pv} \in [0,1]$ for all $p \in \mathrm{Pa}(v)$, then the propagated uncertainty is bounded:
\[
0 \le r_v \le 1.
\]
In particular, if $\mathrm{Pa}(v) = \emptyset$, the uncertainty reduces to the boundary condition $r_v = u_v$.
\end{proposition}

\begin{proposition}[\textbf{Uncertainty Attenuation}]
\label{prop:attenuation}
Let $R_v = \max_{p \in \mathrm{Pa}(v)} r_p$ denote the maximum uncertainty among all parent nodes. If $u_v < R_v$ and 
\[
1 - \prod_{p \in \mathrm{Pa}(v)} (1 - \alpha_{pv}r_p) < \frac{R_v - u_v}{1 - u_v},
\]
then $r_v < R_v$. Therefore, the propagation rule exhibits non-monotonicity, meaning a downstream agent can have lower uncertainty than its parent through low local uncertainty or rigorous critique (i.e., small $\alpha_{pv}$).
\end{proposition}

\begin{proposition}[\textbf{Linear Scalability}]
\label{prop:scalability}
For any MAS execution graph $\mathcal{G}=(\mathcal{V},\mathcal{E})$, the node-wise uncertainties $\{r_v\}_{v \in \mathcal{V}}$ can be computed exactly in a single topological forward pass. The total time complexity scales linearly with the graph size:
\[
\mathcal{O}(|\mathcal{V}| + |\mathcal{E}|).
\]
Consequently, the linear complexity guarantees that Eq.~\eqref{eq:local_recurrence} is computationally efficient, enabling real-time UQ with minimal overhead across arbitrary and dynamic MAS topologies.
\end{proposition}

\section{Empirical Evaluations}

\paragraph{Tasks and Datasets.}
We evaluate \textsc{PropUQ-MAS} across three benchmarks spanning two reasoning-intensive domains:  (i) \textbf{Math and Science Reasoning}, including GSM8K~\cite{gsm8k} and MedQA~\cite{medqa}; and (ii) \textbf{Code Generation}, including MBPP-Plus~\cite{mbppplus}. Detailed descriptions of each benchmark are provided in Appendix~\ref{app:datasets}.


\paragraph{Base Models.}
We evaluate \textsc{PropUQ-MAS} on open-source instruction-tuned models from two families: \texttt{Qwen3-4B}, \texttt{Qwen3-8B}, and \texttt{Qwen3-14B} from the Qwen3 family~\citep{yang2025qwen3}, and \texttt{gemma-3-12b-it}~\citep{gemmateam2025gemma3}. Qwen3 models test robustness across model scales, while Gemma-3 evaluates cross-family generalization.

\paragraph{MAS Structures.}
To evaluate whether \textsc{PropUQ-MAS} generalizes across different interaction topologies, we consider three representative MAS structures: \textit{sequential}~\citep{chain1,chain2}, \textit{hierarchical}~\citep{hierarchical1,hierarchical2}, and \textit{decentralized}~\citep{mesh1,mesh2}. Unless otherwise specified, each MAS contains four agents following the common practice of LLM-MAS studies~\citep{4agents2,4agents1}. This scale provides a controlled yet non-trivial setting, offering sufficient structural expressivity to map each topology while maintaining computational tractability. The evaluated MAS interaction topologies are illustrated in Figure~\ref{fig:mas_structures}, with detailed structure descriptions provided in Appendix~\ref{app:mas_structures}.


\begin{figure}[t]
    \centering
    \includegraphics[width=\linewidth]{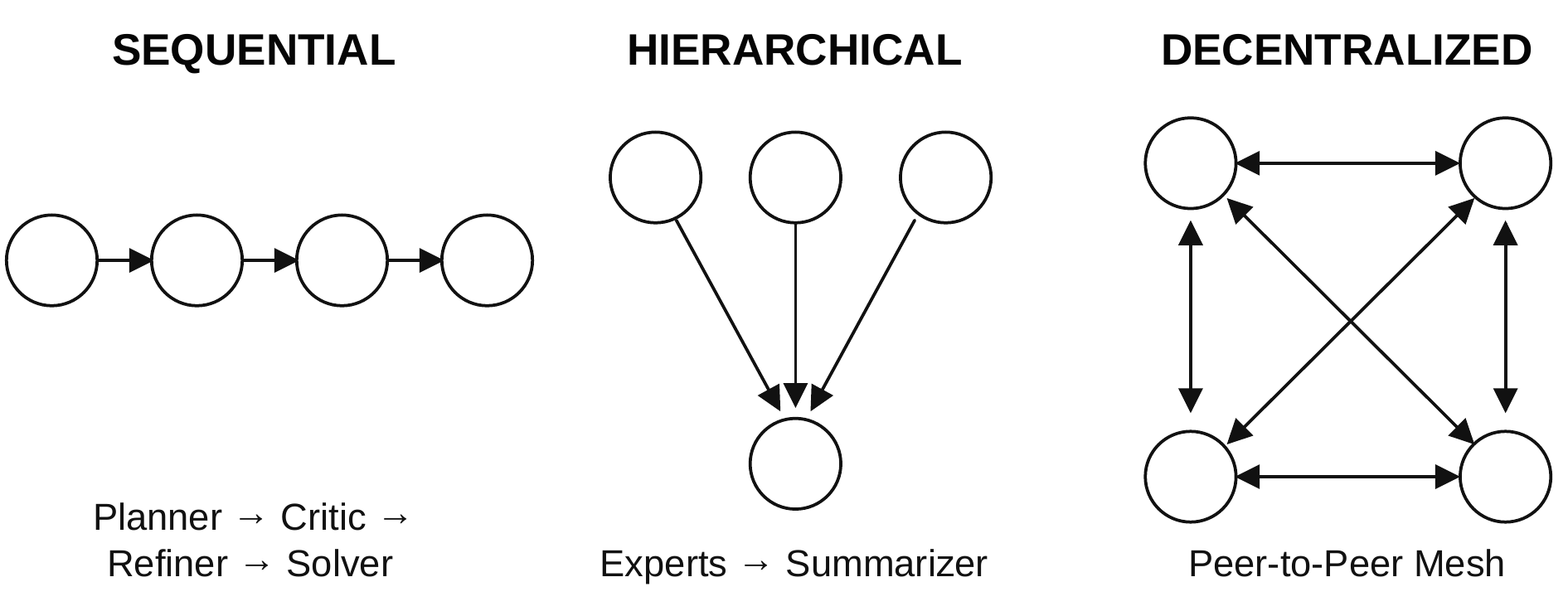} 
    \caption{Illustration of the three evaluated MAS interaction topologies.}
    \label{fig:mas_structures}
    \vspace{-0.2cm}
\end{figure}

\begin{table*}[t]
\centering
\caption{Final-output uncertainty estimation across MAS topologies. \textsc{PropUQ-MAS} is applied on top of each local UQ baseline; shaded cells indicate relative changes over the corresponding baseline, where green/red denotes gains/drops and darker colors indicate larger absolute changes.}
\label{tab:exp1}
\vspace{-0.2cm}
\fontsize{7.5pt}{8.7pt}\selectfont
\begingroup
\setlength{\tabcolsep}{3.5pt}
\renewcommand{\arraystretch}{1.3}
\begin{tabular}{l|l|c c|c c|c c|c c|c c|c c}
\hline
\multirow{2}{*}{\textbf{Dataset}} & \multirow{2}{*}{\textbf{Metric}} & \multicolumn{4}{c}{\textbf{Qwen3-4B}} & \multicolumn{4}{|c}{\textbf{Qwen3-8B}} & \multicolumn{4}{|c}{\textbf{Gemma-3-12B}} \\
\cmidrule(lr){3-6} \cmidrule(lr){7-10} \cmidrule(lr){11-14}
 & & MSP & +\textsc{PropUQ} & Verb. & +\textsc{PropUQ} & MSP & +\textsc{PropUQ} & Verb. & +\textsc{PropUQ} & MSP & +\textsc{PropUQ} & Verb. & +\textsc{PropUQ} \\
\hline
\multicolumn{14}{>{\cellcolor[gray]{0.95}}c}{\textbf{\textit{Sequential MAS}}} \\
\hline
\multirow{2}{*}{\textbf{GSM8K}} & AUROC & 0.578 & \cellcolor[HTML]{DFF2DF}\textbf{0.617} & 0.466 & \cellcolor[HTML]{DCF1DB}\textbf{0.501} & 0.605 & \cellcolor[HTML]{EFF8EF}\textbf{0.625} & 0.505 & \cellcolor[HTML]{B3E0B2}\textbf{0.680} & 0.471 & \cellcolor[HTML]{D2ECD1}\textbf{0.517} & 0.581 & \cellcolor[HTML]{C9E9C8}\textbf{0.651} \\
 & PRR & 0.160 & \cellcolor[HTML]{B3E0B2}\textbf{0.238} & -0.073 & \cellcolor[HTML]{F8DFDF}-0.068 & 0.224 & \cellcolor[HTML]{BEE5BD}\textbf{0.257} & -0.028 & \cellcolor[HTML]{B3E0B2}\textbf{0.340} & 0.005 & \cellcolor[HTML]{B3E0B2}\textbf{0.102} & 0.156 & \cellcolor[HTML]{B3E0B2}\textbf{0.250} \\
\hline
\multirow{2}{*}{\textbf{MBPP+}} & AUROC & 0.679 & \cellcolor[HTML]{E1F3E0}\textbf{0.722} & 0.577 & \cellcolor[HTML]{E6F5E5}\textbf{0.607} & 0.583 & \cellcolor[HTML]{CDEACC}\textbf{0.648} & 0.562 & \cellcolor[HTML]{F2FAF2}\textbf{0.577} & 0.544 & \cellcolor[HTML]{DDF1DD}\textbf{0.583} & 0.684 & \cellcolor[HTML]{D7EFD7}\textbf{0.742} \\
 & PRR & 0.340 & \cellcolor[HTML]{CCEACC}\textbf{0.378} & 0.090 & \cellcolor[HTML]{B3E0B2}\textbf{0.173} & 0.155 & \cellcolor[HTML]{B3E0B2}\textbf{0.282} & 0.064 & \cellcolor[HTML]{B3E0B2}\textbf{0.120} & 0.079 & \cellcolor[HTML]{B3E0B2}\textbf{0.128} & 0.277 & \cellcolor[HTML]{B3E0B2}\textbf{0.465} \\
\hline
\multirow{2}{*}{\textbf{MedQA}} & AUROC & 0.649 & \cellcolor[HTML]{C0E5BF}\textbf{0.742} & 0.428 & \cellcolor[HTML]{B3E0B2}\textbf{0.607} & 0.754 & \cellcolor[HTML]{CAEACA}\textbf{0.842} & 0.600 & \cellcolor[HTML]{B3E0B2}\textbf{0.791} & 0.635 & \cellcolor[HTML]{E4F4E4}\textbf{0.668} & 0.627 & \cellcolor[HTML]{DCF1DB}\textbf{0.674} \\
 & PRR & 0.254 & \cellcolor[HTML]{B3E0B2}\textbf{0.410} & -0.167 & \cellcolor[HTML]{B3E0B2}\textbf{0.236} & 0.496 & \cellcolor[HTML]{B3E0B2}\textbf{0.624} & 0.217 & \cellcolor[HTML]{B3E0B2}\textbf{0.482} & 0.232 & \cellcolor[HTML]{ECF7EC}\textbf{0.240} & 0.237 & \cellcolor[HTML]{B3E0B2}\textbf{0.322} \\
\hline
\multicolumn{14}{>{\cellcolor[gray]{0.95}}c}{\textbf{\textit{Hierarchical MAS}}} \\
\hline
\multirow{2}{*}{\textbf{GSM8K}} & AUROC & 0.564 & \cellcolor[HTML]{FBEDED}0.541 & 0.535 & \cellcolor[HTML]{CBEACA}\textbf{0.597} & 0.664 & \cellcolor[HTML]{D9F0D9}\textbf{0.717} & 0.516 & \cellcolor[HTML]{CAE9C9}\textbf{0.577} & 0.535 & \cellcolor[HTML]{F9E3E3}0.498 & 0.549 & \cellcolor[HTML]{D0ECCF}\textbf{0.606} \\
 & PRR & 0.100 & \cellcolor[HTML]{F9E4E4}0.094 & 0.082 & \cellcolor[HTML]{B3E0B2}\textbf{0.222} & 0.321 & \cellcolor[HTML]{B3E0B2}\textbf{0.446} & -0.001 & \cellcolor[HTML]{B3E0B2}\textbf{0.153} & 0.063 & \cellcolor[HTML]{FCEFEF}0.060 & 0.080 & \cellcolor[HTML]{B3E0B2}\textbf{0.204} \\
\hline
\multirow{2}{*}{\textbf{MBPP+}} & AUROC & 0.688 & \cellcolor[HTML]{EAF6EA}\textbf{0.718} & 0.632 & \cellcolor[HTML]{E9F6E9}\textbf{0.661} & 0.619 & \cellcolor[HTML]{B9E2B8}\textbf{0.719} & 0.580 & \cellcolor[HTML]{D6EED6}\textbf{0.630} & 0.630 & \cellcolor[HTML]{DBF0DB}\textbf{0.674} & 0.682 & \cellcolor[HTML]{DEF2DD}\textbf{0.730} \\
 & PRR & 0.335 & \cellcolor[HTML]{B3E0B2}\textbf{0.432} & 0.201 & \cellcolor[HTML]{B3E0B2}\textbf{0.289} & 0.172 & \cellcolor[HTML]{B3E0B2}\textbf{0.401} & 0.081 & \cellcolor[HTML]{B3E0B2}\textbf{0.245} & 0.204 & \cellcolor[HTML]{F7FCF7}\textbf{0.207} & 0.382 & \cellcolor[HTML]{CDEBCD}\textbf{0.423} \\
\hline
\multirow{2}{*}{\textbf{MedQA}} & AUROC & 0.628 & \cellcolor[HTML]{B3E0B2}\textbf{0.760} & 0.621 & \cellcolor[HTML]{C2E6C2}\textbf{0.705} & 0.658 & \cellcolor[HTML]{D9EFD8}\textbf{0.711} & 0.604 & \cellcolor[HTML]{B3E0B2}\textbf{0.722} & 0.567 & \cellcolor[HTML]{E0F2DF}\textbf{0.604} & 0.660 & \cellcolor[HTML]{F8FCF8}\textbf{0.669} \\
 & PRR & 0.238 & \cellcolor[HTML]{B3E0B2}\textbf{0.417} & 0.216 & \cellcolor[HTML]{B3E0B2}\textbf{0.296} & 0.277 & \cellcolor[HTML]{B3E0B2}\textbf{0.349} & 0.223 & \cellcolor[HTML]{B3E0B2}\textbf{0.362} & 0.115 & \cellcolor[HTML]{B3E0B2}\textbf{0.184} & 0.358 & \cellcolor[HTML]{FDF7F7}0.351 \\
\hline
\multicolumn{14}{>{\cellcolor[gray]{0.95}}c}{\textbf{\textit{Decentralized MAS}}} \\
\hline
\multirow{2}{*}{\textbf{GSM8K}} & AUROC & 0.657 & \cellcolor[HTML]{DEF1DD}\textbf{0.704} & 0.544 & \cellcolor[HTML]{E4F4E4}\textbf{0.574} & 0.728 & \cellcolor[HTML]{E5F4E4}\textbf{0.768} & 0.486 & \cellcolor[HTML]{BAE3B9}\textbf{0.563} & 0.563 & \cellcolor[HTML]{FCFEFC}\textbf{0.567} & 0.634 & \cellcolor[HTML]{E7F5E7}\textbf{0.665} \\
 & PRR & 0.311 & \cellcolor[HTML]{B3E0B2}\textbf{0.428} & 0.128 & \cellcolor[HTML]{B3E0B2}\textbf{0.185} & 0.445 & \cellcolor[HTML]{B3E0B2}\textbf{0.534} & -0.004 & \cellcolor[HTML]{B3E0B2}\textbf{0.181} & 0.128 & \cellcolor[HTML]{B3E0B2}\textbf{0.162} & 0.271 & \cellcolor[HTML]{D9EFD8}\textbf{0.294} \\
\hline
\multirow{2}{*}{\textbf{MBPP+}} & AUROC & 0.695 & \cellcolor[HTML]{F4FBF4}\textbf{0.710} & 0.588 & \cellcolor[HTML]{F2FAF1}\textbf{0.604} & 0.701 & \cellcolor[HTML]{EEF8ED}\textbf{0.726} & 0.511 & \cellcolor[HTML]{C1E6C0}\textbf{0.583} & 0.557 & \cellcolor[HTML]{BAE3B9}\textbf{0.645} & 0.670 & \cellcolor[HTML]{DCF1DC}\textbf{0.720} \\
 & PRR & 0.382 & \cellcolor[HTML]{CAEACA}\textbf{0.426} & 0.140 & \cellcolor[HTML]{B3E0B2}\textbf{0.210} & 0.332 & \cellcolor[HTML]{CEEBCD}\textbf{0.367} & 0.014 & \cellcolor[HTML]{B3E0B2}\textbf{0.117} & 0.096 & \cellcolor[HTML]{B3E0B2}\textbf{0.262} & 0.342 & \cellcolor[HTML]{BDE4BD}\textbf{0.393} \\
\hline
\multirow{2}{*}{\textbf{MedQA}} & AUROC & 0.749 & \cellcolor[HTML]{D1ECD0}\textbf{0.824} & 0.693 & \cellcolor[HTML]{E4F4E4}\textbf{0.731} & 0.812 & \cellcolor[HTML]{FAFDFA}\textbf{0.821} & 0.718 & \cellcolor[HTML]{B3E0B2}\textbf{0.854} & 0.599 & \cellcolor[HTML]{E1F3E1}\textbf{0.636} & 0.576 & \cellcolor[HTML]{E2F3E2}\textbf{0.611} \\
 & PRR & 0.416 & \cellcolor[HTML]{B3E0B2}\textbf{0.563} & 0.355 & \cellcolor[HTML]{FAEAEA}0.337 & 0.539 & \cellcolor[HTML]{E6F5E6}\textbf{0.568} & 0.429 & \cellcolor[HTML]{B3E0B2}\textbf{0.626} & 0.155 & \cellcolor[HTML]{B3E0B2}\textbf{0.208} & 0.160 & \cellcolor[HTML]{B3E0B2}\textbf{0.200} \\
\hline
\end{tabular}
\endgroup
\end{table*}

\paragraph{Evaluation Metrics.}
We evaluate uncertainty quality using AUROC and prediction rejection ratio (PRR). 
\yifan{For each node output, we convert its correctness annotation into a binary error label, setting the error label to 1 if the output is incorrect and 0 otherwise.}
AUROC measures the ranking quality of uncertainty scores, testing whether incorrect outputs receive higher uncertainty than correct ones. In contrast, PRR evaluates the usefulness of uncertainty for selective prediction: outputs are rejected from most to least uncertain, and a better estimator should remove errors earlier, leading to higher retained accuracy. Thus, AUROC measures threshold-free error ranking, while PRR measures decision-level utility under uncertainty-based rejection. Detailed definitions are provided in Appendix~\ref{app:metrics}. Higher AUROC and PRR indicate better uncertainty estimation. 


\paragraph{Baselines.}
To assess compatibility with different single-agent UQ methods, we select two representative local uncertainty estimators for LLMs: 
verbalized self-evaluation (Verb.)~\cite{ask4conf1} and
Maximum Sequence Probability (MSP)~\cite{lm-polygraph} for token-level predictive probability. 
Each estimator is evaluated as a standalone baseline and is also used to provide the local uncertainty $u_v$ for \textsc{PropUQ-MAS}.
Because \textsc{PropUQ-MAS} operates with a single MAS forward pass, we treat sampling-based UQ methods that require multiple sampled executions as complementary rather than local UQ baselines. We nevertheless compare with concurrent sampling-based methods, including MATU~\cite{chen2026every} and UProp~\cite{duan2025uprop}, in Appendix~\ref{app:concurrent_sampling_baselines}.

\paragraph{Implementation Details.}
All main experiments are run on NVIDIA GH200 120GB GPUs. Unless otherwise specified, generation uses temperature 0.6, top-$p$ 0.95, and repetition penalty 1.05. The maximum generation length is set to 8192 tokens for math and science reasoning tasks and 32768 tokens for code-generation tasks. 

\begin{table*}[t]
\centering
\caption{Intermediate-agent uncertainty estimation under decentralized MAS. \textsc{PropUQ} cells are shaded by relative change over each baseline: green/red indicates gains/drops, and darker colors indicate larger absolute changes.}
\label{tab:exp2-agent}
\vspace{-0.2cm}
\fontsize{7.2pt}{8.3pt}\selectfont
\begingroup
\setlength{\tabcolsep}{3.1pt}
\renewcommand{\arraystretch}{1.3}
\begin{tabular}{l|l|l|c c|c c|c c|c c|c c|c c}
\hline
\multirow{2}{*}{\textbf{Dataset}} & \multirow{2}{*}{\textbf{Agent}} & \multirow{2}{*}{\textbf{Metric}} & \multicolumn{4}{c}{\textbf{Qwen3-4B}} & \multicolumn{4}{|c}{\textbf{Qwen3-8B}} & \multicolumn{4}{|c}{\textbf{Gemma-3-12B}} \\
\cmidrule(lr){4-7} \cmidrule(lr){8-11} \cmidrule(lr){12-15}
 & & & MSP & +\textsc{PropUQ} & Verb. & +\textsc{PropUQ} & MSP & +\textsc{PropUQ} & Verb. & +\textsc{PropUQ} & MSP & +\textsc{PropUQ} & Verb. & +\textsc{PropUQ} \\
\hline
\multirow[c]{4}{*}{\textbf{GSM8K}} & \multirow{2}{*}{Agent 2} & AUROC & 0.667 & \cellcolor[HTML]{EBF7EA}\textbf{0.695} & 0.586 & \cellcolor[HTML]{E2F3E1}\textbf{0.622} & 0.754 & \cellcolor[HTML]{F5FBF5}\textbf{0.769} & 0.544 & \cellcolor[HTML]{CFEBCE}\textbf{0.601} & 0.589 & \cellcolor[HTML]{FDFEFD}\textbf{0.591} & 0.640 & \cellcolor[HTML]{EDF8EC}\textbf{0.664} \\
 &  & PRR & 0.372 & \cellcolor[HTML]{C4E7C3}\textbf{0.421} & 0.144 & \cellcolor[HTML]{B3E0B2}\textbf{0.224} & 0.499 & \cellcolor[HTML]{E3F4E3}\textbf{0.528} & 0.056 & \cellcolor[HTML]{B3E0B2}\textbf{0.201} & 0.185 & 0.184 & 0.248 & \cellcolor[HTML]{BFE5BE}\textbf{0.284} \\
\cline{2-15}
 & \multirow{2}{*}{Agent 3} & AUROC & 0.668 & \cellcolor[HTML]{EDF8ED}\textbf{0.692} & 0.549 & \cellcolor[HTML]{D3EDD3}\textbf{0.601} & 0.724 & \cellcolor[HTML]{E7F5E7}\textbf{0.760} & 0.556 & \cellcolor[HTML]{C7E8C6}\textbf{0.625} & 0.561 & \cellcolor[HTML]{FBFDFB}\textbf{0.566} & 0.629 & \cellcolor[HTML]{E5F4E5}\textbf{0.663} \\
 &  & PRR & 0.369 & \cellcolor[HTML]{C0E5BF}\textbf{0.421} & 0.120 & \cellcolor[HTML]{B3E0B2}\textbf{0.221} & 0.416 & \cellcolor[HTML]{B3E0B2}\textbf{0.509} & 0.084 & \cellcolor[HTML]{B3E0B2}\textbf{0.245} & 0.125 & \cellcolor[HTML]{B3E0B2}\textbf{0.151} & 0.257 & \cellcolor[HTML]{DBF0DA}\textbf{0.276} \\
\hline
\multirow[c]{4}{*}{\textbf{MBPP+}} & \multirow{2}{*}{Agent 2} & AUROC & 0.692 & \cellcolor[HTML]{E2F3E1}\textbf{0.734} & 0.569 & \cellcolor[HTML]{EDF8ED}\textbf{0.590} & 0.690 & \cellcolor[HTML]{E5F4E4}\textbf{0.729} & 0.490 & \cellcolor[HTML]{CBEACA}\textbf{0.546} & 0.687 & \cellcolor[HTML]{F9E4E4}0.642 & 0.666 & \cellcolor[HTML]{EDF7EC}\textbf{0.691} \\
 &  & PRR & 0.368 & \cellcolor[HTML]{B3E0B2}\textbf{0.454} & 0.126 & \cellcolor[HTML]{B3E0B2}\textbf{0.169} & 0.325 & \cellcolor[HTML]{BEE4BD}\textbf{0.373} & -0.026 & \cellcolor[HTML]{B3E0B2}\textbf{0.098} & 0.355 & \cellcolor[HTML]{FFFEFE}0.353 & 0.361 & \cellcolor[HTML]{D3EDD3}\textbf{0.395} \\
\cline{2-15}
 & \multirow{2}{*}{Agent 3} & AUROC & 0.743 & \cellcolor[HTML]{F3FAF3}\textbf{0.761} & 0.627 & \cellcolor[HTML]{FDFEFD}\textbf{0.630} & 0.699 & \cellcolor[HTML]{FFFFFF}0.698 & 0.566 & \cellcolor[HTML]{E1F3E1}\textbf{0.602} & 0.639 & \cellcolor[HTML]{EBF7EB}\textbf{0.665} & 0.663 & \cellcolor[HTML]{E1F3E1}\textbf{0.705} \\
 &  & PRR & 0.415 & \cellcolor[HTML]{B7E2B7}\textbf{0.484} & 0.253 & \cellcolor[HTML]{FFFDFD}0.252 & 0.349 & \cellcolor[HTML]{F7FCF7}\textbf{0.355} & 0.158 & \cellcolor[HTML]{B3E0B2}\textbf{0.190} & 0.241 & \cellcolor[HTML]{B3E0B2}\textbf{0.290} & 0.339 & \cellcolor[HTML]{B7E2B6}\textbf{0.395} \\
\hline
\multirow[c]{4}{*}{\textbf{MedQA}} & \multirow{2}{*}{Agent 2} & AUROC & 0.770 & \cellcolor[HTML]{ECF7EC}\textbf{0.799} & 0.715 & \cellcolor[HTML]{E6F5E5}\textbf{0.752} & 0.796 & \cellcolor[HTML]{F7FCF6}\textbf{0.810} & 0.667 & \cellcolor[HTML]{B3E0B2}\textbf{0.816} & 0.656 & \cellcolor[HTML]{F7FCF7}\textbf{0.667} & 0.622 & \cellcolor[HTML]{FDFEFD}\textbf{0.625} \\
 &  & PRR & 0.481 & \cellcolor[HTML]{DFF2DF}\textbf{0.513} & 0.408 & \cellcolor[HTML]{EDF8ED}\textbf{0.423} & 0.477 & \cellcolor[HTML]{DFF2DF}\textbf{0.509} & 0.338 & \cellcolor[HTML]{B3E0B2}\textbf{0.509} & 0.284 & \cellcolor[HTML]{E6F5E5}\textbf{0.298} & 0.297 & \cellcolor[HTML]{FDF4F4}0.290 \\
\cline{2-15}
 & \multirow{2}{*}{Agent 3} & AUROC & 0.781 & \cellcolor[HTML]{ECF7EC}\textbf{0.812} & 0.629 & \cellcolor[HTML]{CBEACA}\textbf{0.701} & 0.776 & \cellcolor[HTML]{E3F4E2}\textbf{0.822} & 0.695 & \cellcolor[HTML]{B3E0B2}\textbf{0.824} & 0.581 & \cellcolor[HTML]{D4EED4}\textbf{0.634} & 0.553 & \cellcolor[HTML]{D0ECCF}\textbf{0.610} \\
 &  & PRR & 0.479 & \cellcolor[HTML]{BBE3BA}\textbf{0.554} & 0.239 & \cellcolor[HTML]{B3E0B2}\textbf{0.285} & 0.459 & \cellcolor[HTML]{B3E0B2}\textbf{0.539} & 0.390 & \cellcolor[HTML]{B3E0B2}\textbf{0.562} & 0.174 & \cellcolor[HTML]{B3E0B2}\textbf{0.212} & 0.169 & \cellcolor[HTML]{B9E3B9}\textbf{0.196} \\
\hline
\end{tabular}
\endgroup
\end{table*}

\begin{table}[t]
\centering
\caption{Intermediate-agent uncertainty estimation under sequential MAS on MedQA with Qwen3-8B, using GPT-5.5 judgments.}
\label{tab:sequential_node_uq}
\vspace{-0.1cm}
\small
\renewcommand{\arraystretch}{1.2}
\begin{tabular}{l|l|cc}
\hline
\textbf{Agent} & \textbf{UQ Method} & \textbf{AUROC} & \textbf{PRR} \\
\hline
\multirow{2}{*}{Critic} & Verb. & 0.7516 & 0.4264 \\
 & +\textsc{PropUQ} & \textbf{0.7872} & \textbf{0.5779} \\
 \hline
\multirow{2}{*}{Refiner} & Verb. & 0.6709 & 0.3869 \\
 & +\textsc{PropUQ} & \textbf{0.7886} & \textbf{0.5012} \\
\hline
\end{tabular}
\end{table}

\subsection{Final-Output UQ Performance}
\label{sec:final_output_uq}

We first evaluate whether \textsc{PropUQ-MAS} improves uncertainty estimation for the final answer produced by a MAS. For each instance, we compare the uncertainty score of the final output produced by a standalone local UQ baseline with the propagated uncertainty score computed by \textsc{PropUQ-MAS}.  This comparison tests whether modeling uncertainty propagation better detects final-answer errors.

\yifan{Table~\ref{tab:exp1} shows that \textsc{PropUQ-MAS} improves final-output UQ in most settings, with median relative gains of $+7.32\%$ in AUROC and $+41.36\%$ in PRR across datasets, model families, and MAS topologies. The gains are especially clear when the local UQ baseline is weak or noisy, suggesting that final-agent confidence alone can miss errors inherited from upstream agents. By incorporating parent uncertainty and edge-level acceptance, \textsc{PropUQ-MAS} captures this interaction-induced risk and yields more informative final-answer uncertainty.}
\yifan{The results hold across model families and scales, suggesting that \textsc{PropUQ-MAS} works as a training-free, model-agnostic propagation layer rather than a model-specific calibration method. The overall performance gain supports the importance of explicitly modeling uncertainty propagation in MAS reliability assessment. Additional results on \texttt{Qwen3-14B} are provided in Appendix~\ref{app:additional_experiments}.}

\subsection{Intermediate-Step UQ Performance}
\label{sec:node_level_uq}

We further evaluate whether \textsc{PropUQ-MAS} provides reliable uncertainty signals for intermediate MAS outputs, rather than only for the final answer. This experiment tests a key requirement of MAS reliability assessment: the capacity to pinpoint early-stage failures that precipitate final system errors. 

\subsubsection{Verifiable Intermediate Steps}
We first conduct quantitative analysis within the decentralized MAS setting, where each intermediate agent output is a complete candidate solution and can be checked against the ground-truth label. 
This configuration enables a direct intermediate-step evaluation of uncertainty scores. 
We compare \textsc{PropUQ-MAS} with local UQ baselines to measure the benefit of modeling interaction-aware propagation at each step. 

\yifan{Table~\ref{tab:exp2-agent} reports intermediate-agent UQ results, 
with node-averaged results in Appendix Table~\ref{tab:exp2-mean}.} We omit Agent~1 because it has no incoming messages, so its propagated uncertainty is identical to its local uncertainty, i.e., $r_v=u_v$. We also omit Agent~4 in the intermediate-agent table because its output corresponds to the final MAS answer, whose results are already reported in Table~\ref{tab:exp1}. Across the evaluated intermediate agents, \textsc{PropUQ-MAS} achieves average relative gains of $+6.10\%$ in AUROC and $+47.58\%$ in PRR, showing that modeling propagated risk helps identify unreliable intermediate outputs, not only unreliable final answers. The PRR gains indicate that high-risk intermediate steps can be prioritized for rejection or intervention before they affect later agents.


\begin{figure*}[htbp]
    \centering
    \includegraphics[width=\linewidth]{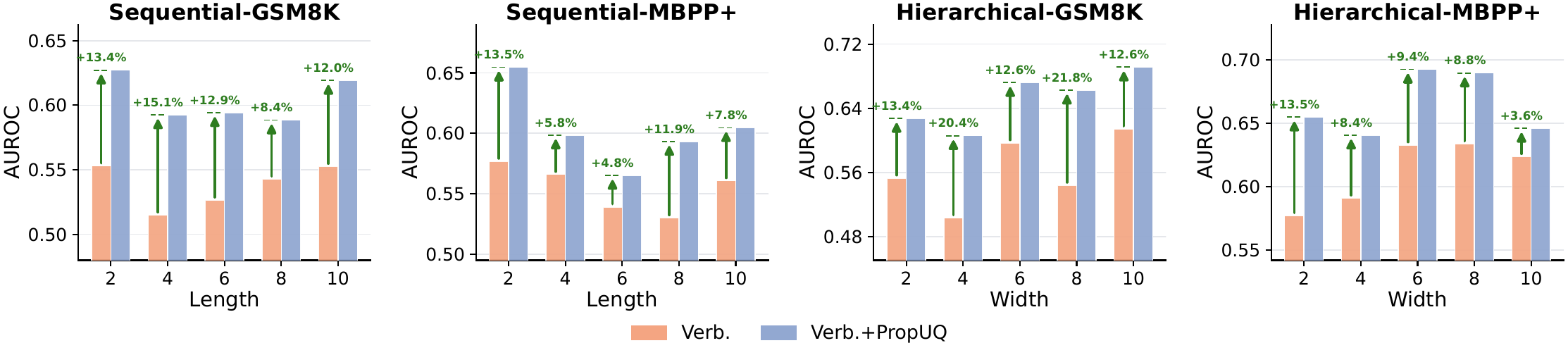} 
    \vspace{-0.7cm}
    \caption{
    Structural scale generalization results measured by AUROC. We evaluate the robustness of uncertainty propagation across varying sequential chain lengths and hierarchical expert-layer widths.
    }
    \label{fig:generalization_auroc}
\end{figure*}

\subsubsection{LLM-Judged Reasoning Steps}

For intermediate reasoning steps without direct ground-truth labels, we conduct an LLM-judged evaluation on 300 randomly sampled MedQA instances under the sequential MAS setting using Qwen3-8B.
We use GPT-5.5 as an LLM judge~\citep{zheng2023judging} to annotate Critic and Refiner outputs as erroneous when they contain substantive medical or reasoning errors likely to mislead downstream agents; the judge prompt is provided in Appendix~\ref{app:node_judge_prompt}. 
We exclude the Planner because it is a source node with no upstream inputs and therefore does not involve uncertainty propagation. For hierarchical MAS, the summarizer is the only node that aggregates upstream uncertainty, and its output is the final answer already evaluated in Table~\ref{tab:exp1}. The expert agents are source nodes, leaving no additional propagated intermediate node for evaluation.

Table~\ref{tab:sequential_node_uq} shows consistent improvements from \textsc{PropUQ-MAS} for both intermediate agents. On 50 randomly selected examples, human annotations agree with the GPT-5.5 judgments in $92.0\%$ of cases, supporting the reliability of the judge-derived labels. These results show that propagation-aware UQ extends beyond complete candidate answers to intermediate reasoning states.

Overall, these results support \textsc{PropUQ-MAS} as an online reliability monitor for MAS execution. By attaching propagated uncertainty to intermediate steps, it can identify early-stage errors and support interventions such as localized replanning, external verification, additional critique, or early stopping before these errors affect the final answer. Additional results on \texttt{Qwen3-14B} are provided in Appendix~\ref{app:additional_experiments}.




\begin{table}[t]
\centering
\caption{
Ablation on self-reported acceptance weights using Qwen3-8B. The w/o $\hat{\alpha}$ variant fixes all edge weights to one, while w/ $\hat{\alpha}$ uses self-reported acceptance scores. Bold marks the better value in each pair.
}
\vspace{-0.2cm}
\label{tab:exp4-ask4conf-acceptance-ablation}
\fontsize{7.5pt}{8.7pt}\selectfont
\begingroup
\setlength{\tabcolsep}{4.0pt}
\renewcommand{\arraystretch}{1.35}
\begin{tabular}{l|l|c c|c c|c c}
\hline
\multirow{2}{*}{\textbf{Dataset}} & \multirow{2}{*}{\textbf{Metric}} & \multicolumn{2}{c}{\textbf{Sequential}} & \multicolumn{2}{|c}{\textbf{Hierarchical}} & \multicolumn{2}{|c}{\textbf{Decentralized}} \\
\cmidrule(lr){3-4} \cmidrule(lr){5-6} \cmidrule(lr){7-8}
 & & w/o $\hat{\alpha}$ & w/ $\hat{\alpha}$ & w/o $\hat{\alpha}$ & w/ $\hat{\alpha}$ & w/o $\hat{\alpha}$ & w/ $\hat{\alpha}$ \\
\hline
\multirow{2}{*}{\textbf{GSM8K}} & AUROC & 0.672 & \textbf{0.680} & 0.569 & \textbf{0.577} & 0.519 & \textbf{0.563} \\
 & PRR & 0.331 & \textbf{0.340} & 0.148 & \textbf{0.153} & 0.142 & \textbf{0.181} \\
\hline
\multirow{2}{*}{\textbf{MBPP+}} & AUROC & 0.573 & \textbf{0.577} & 0.615 & \textbf{0.630} & 0.576 & \textbf{0.583} \\
 & PRR & 0.114 & \textbf{0.120} & 0.239 & \textbf{0.245} & 0.093 & \textbf{0.117} \\
\hline
\multirow{2}{*}{\textbf{MedQA}} & AUROC & 0.785 & \textbf{0.791} & 0.719 & \textbf{0.722} & 0.848 & \textbf{0.854} \\
 & PRR & 0.476 & \textbf{0.482} & 0.345 & \textbf{0.362} & 0.612 & \textbf{0.626} \\
\hline
\end{tabular}
\endgroup
\end{table}

\subsection{MAS Scale Analysis}
\label{sec:generalization}

\yifan{We further examine whether \textsc{PropUQ-MAS} remains effective as MAS scale changes. Following prior work on MAS scaling~\cite{scaling}, we vary two factors that directly affect uncertainty propagation: the chain length in sequential systems and the number of parent inputs in hierarchical systems. Each agent acts as both a critic and a solver, keeping agent behavior comparable across scales while preserving reasoning capability. Prompts are provided in Appendix~\ref{app:prompts}.} 

Figure~\ref{fig:generalization_auroc} reports AUROC results on GSM8K and MBPP-Plus, with additional PRR results in Appendix~\ref{app:additional_experiments}. \textsc{PropUQ-MAS} consistently improves over the local verbalized UQ baseline across all tested sequential lengths and hierarchical widths. These results indicate that the propagation rule can track uncertainty over longer chains and aggregate risks from larger parent sets, supporting its scalability under different MAS sizes.

\subsection{Ablation Study}
\label{sec:ablation}

We conduct an ablation study to examine whether self-reported acceptance scores provide useful edge-level signals for uncertainty propagation.  The ablated variant, w/o $\hat{\alpha}$, sets all edge weights to 1, uniformly propagating parent uncertainty regardless of downstream acceptance or critique. The full model, w/ $\hat{\alpha}$, uses the self-reported acceptance score as the plug-in transmission weight for each edge. Table~\ref{tab:exp4-ask4conf-acceptance-ablation} shows that w/ $\hat{\alpha}$ consistently improves over w/o $\hat{\alpha}$ on both AUROC and PRR across three MAS topologies with Qwen3-8B, indicating that self-reported $\hat{\alpha}$ provides useful interaction signals beyond uniform uncertainty propagation. The results support the correction-sensitive design of \textsc{PropUQ-MAS}, where accepted upstream risks contribute more than critiqued ones.



\subsection{Validation of Self-Reported Edge Weights}
\label{sec:edge_weight_calibration}

The error-conditioned edge parameter $\alpha_{pv}=\Pr(M_{pv}=A\mid E_p=1)$ is not directly accessible during online execution, as it depends on the correctness of the parent output. We therefore examine whether the self-reported proxy $\hat{\alpha}_{pv}$ preserves the relative strength of error acceptance. Under the decentralized MAS topology, we conduct an edge-level analysis across three datasets and four models. Specifically, we isolate execution edges whose parent is incorrect ($E_p=1$), bin them by the child node's self-reported adoption score $\hat{\alpha}_{pv}$, and measure the error-inheritance rate, i.e., how often the child duplicates the parent's error.

\begin{figure}[t]
    \centering
    \includegraphics[width=\linewidth]{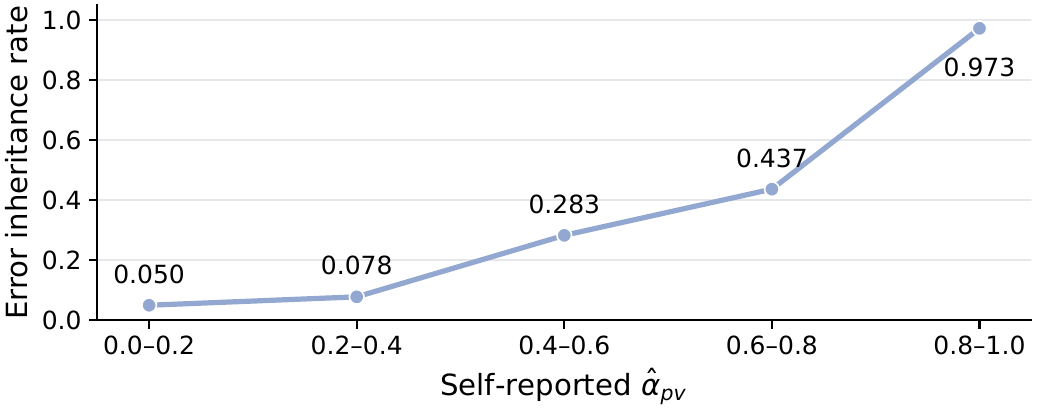}
    \vspace{-0.6cm}
    \caption{Child error-inheritance rate across adoption-score bins for incorrect-parent cases, aggregated over three datasets and four models.}
    \label{fig:edge_weight_relative_calibration}
\end{figure}

Figure~\ref{fig:edge_weight_relative_calibration} shows a monotonic alignment between the self-reported score and the empirical error-inheritance rate, increasing from $0.050$ in the lowest bin to $0.973$ in the highest. This monotonicity demonstrates that the self-reported $\hat{\alpha}_{pv}$ serves as a reliable proxy for error-conditioned acceptance strength. Crucially, it provides this signal online without additional training or inference passes. 
More generally, the \textsc{PropUQ-MAS} propagation recurrence in Eq.~\eqref{eq:local_recurrence} is agnostic to the specific choice of edge-weight estimator. When labeled calibration data are available, the self-reported $\hat{\alpha}{pv}$ can therefore be replaced by calibrated or learned estimators that more accurately approximate the error-conditioned edge parameter $\alpha_{pv}$.

\begin{figure}[t]
\centering
\begin{tcolorbox}[
    width=\linewidth,
    colback=gray!4!white,
    colframe=gray!70!black,
    colbacktitle=gray!70!black,
    coltitle=white,
    title=Case Study from GSM8K under Hierarchical MAS,
    fonttitle=\bfseries\footnotesize,
    left=5pt, right=5pt, top=4pt, bottom=4pt,
    boxsep=2pt,
    arc=2pt
]
\footnotesize
\setlength{\parskip}{2pt}
\setlength{\parindent}{0pt}

\textbf{Question.}
A team of 4 painters worked on a mansion for \(3/8\) of a day every day for 3 weeks.
How many hours of work did each painter put in?

\textbf{Gold Answer:} \textcolor{ForestGreen}{189}
\hfill
\textbf{Final Prediction:} \textcolor{BrickRed}{47.25 (incorrect)}

\noindent\hdashrule{\linewidth}{0.4pt}{2pt 2pt}

\vspace{2pt}
\textbf{Math Agent.}
\textcolor{BrickRed}{
\textit{Answer:} compute 3 weeks as 21 days and \(3/8\) of a day as 9 hours, giving \(21\times9=189\) hours. Then divide by 4 painters, concluding \(189/4=47.25\) hours per painter.
}

\textit{Local uncertainty:} 0.20\\
\textit{Propagation-aware uncertainty:} 0.20\\
\textit{Acceptance by Summarizer:} 0.90

\vspace{1pt}
\noindent\hdashrule{\linewidth}{0.4pt}{2pt 2pt}

\textbf{Science Agent.}
\textcolor{BrickRed}{
\textit{Answer:} treat \(21\times 3/8=63/8\) days as total team time, divide by 4 painters to get \(63/32\) days per painter, and convert to hours: \((63/32)\times24=47.25\).
}

\textit{Local uncertainty:} 0.30\\
\textit{Propagation-aware uncertainty:} 0.30\\
\textit{Acceptance by Summarizer:} 0.90

\vspace{1pt}
\noindent\hdashrule{\linewidth}{0.4pt}{2pt 2pt}

\textbf{Code Agent.}
\textcolor{ForestGreen}{
\textit{Answer:} each painter works \(3/8\) of a day, i.e., \((3/8)\times24=9\) hours per day. Over 3 weeks, this is 21 days, so each painter works \(9\times21=189\) hours.
}

\textit{Local uncertainty:} 0.05\\
\textit{Propagation-aware uncertainty:} 0.05\\
\textit{Acceptance by Summarizer:} 0.30

\vspace{1pt}
\noindent\hdashrule{\linewidth}{0.4pt}{2pt 2pt}

\textbf{Summarizer.}
\textcolor{BrickRed}{
\textit{I follow the Math and Science agents:} first compute 21 days and 9 hours per day, giving 189 hours. Since there are 4 painters, divide by 4 and output 47.25. The Code Agent's answer 189 is treated as total team hours rather than per-painter hours.
}

\textit{Local uncertainty:} 0.10\\
\textit{Propagation-aware uncertainty:} 0.469

\end{tcolorbox}
\vspace{-0.3cm}
\caption{
Case study from GSM8K under the hierarchical MAS topology. The summarizer follows erroneous agents with low local uncertainty, while \textsc{PropUQ-MAS} captures the propagated risk.
}
\label{fig:case_study_hierarchical_line1047}
\vspace{-0.2cm}
\end{figure}

\subsection{Case Study}
\label{sec:case_study}

Figure~\ref{fig:case_study_hierarchical_line1047} presents a GSM8K case under the hierarchical MAS topology. The summarizer is locally confident but follows two erroneous upstream agents while assigning low acceptance to the correct one, leading to an incorrect final answer. This illustrates a common MAS failure mode: local confidence can underestimate risk when downstream agents selectively rely on unreliable upstream information. By combining parent uncertainty with edge-level acceptance, \textsc{PropUQ-MAS} assigns higher propagation-aware uncertainty to the final output, making the propagated error more detectable. A complementary sequential MAS case study is provided in Appendix~\ref{app:sequential_case_study}.

\section{Conclusion}

We introduce \textsc{PropUQ-MAS}, a propagation-aware uncertainty quantification framework for LLM-based multi-agent systems. Unlike standard UQ methods that estimate the reliability of isolated responses, \textsc{PropUQ-MAS} models MAS execution as a communication-structured graph and estimates node-wise reliability by combining intrinsic local uncertainty with cascading uncertainty inherited from upstream messages. Its recursive propagation rule enables efficient uncertainty tracking in a single topological forward pass without additional training. Experiments across reasoning and code-generation tasks show consistent improvements over single-agent UQ baselines for both final-answer and intermediate-step UQ, with gains across topologies and backbone models.

\section*{Limitations}

Despite its empirical effectiveness, \textsc{PropUQ-MAS} has several limitations. 

First, the theoretical propagation rule uses the latent error-conditioned transmission parameter $\alpha_{pv}$, while our implementation instantiates it with a self-reported plug-in estimate $\hat{\alpha}_{pv}$. Since the parent error event is unobserved during online execution, this self-reported score may be imperfectly calibrated as a probability. We therefore view $\hat{\alpha}_{pv}$ as an online proxy for edge-level transmission strength rather than a direct observation of the true error-conditioned parameter. Future work could replace this proxy with calibrated, verifier-based, or learned transmission estimators.

Second, our propagation rule relies on edge-wise conditional independence and local-propagation independence to obtain a tractable factorization. These assumptions may be violated when upstream messages jointly affect downstream reasoning or when misleading context simultaneously increases local generation error and acceptance behavior, such as under severe prompt injection or context overload. Despite these approximations, the factorized rule enables efficient single-pass uncertainty estimation and empirically improves over local UQ baselines across tasks, models, and MAS topologies. Future work could relax these assumptions by modeling correlations among local errors, acceptance behavior, and propagated uncertainty.

Third, \textsc{PropUQ-MAS} depends on the quality of the local UQ estimator at each node. While the framework is agnostic to the specific local UQ estimator and can readily incorporate calibrated alternatives, such as temperature scaling or isotonic regression, these methods typically require labeled calibration data and may be less practical for online MAS execution. We therefore use lightweight, single-pass MSP and verbalized uncertainty to support real-time inference. Incorporating stronger local UQ estimators is complementary to our propagation framework and may further improve propagation-aware uncertainty estimation.



\bibliography{custom}

\clearpage
\appendix
\section{Acceptance Probability Elicitation}
\label{app:self_reporting}

We instantiate the edge-level transmission parameter $\alpha_{pv}$ using a training-free self-reporting procedure. In the probabilistic formulation, $\alpha_{pv}$ denotes the error-conditioned acceptance probability, i.e., the probability that node $v$ accepts information from parent node $p$ when the parent output is erroneous. During online MAS execution, this true error-conditioned quantity is not directly observable, since the parent error variable $E_p$ is unknown. We therefore use a self-reported adoption score as a plug-in estimate of the edge-level transmission strength.

Specifically, after node $v$ is generated, the corresponding agent is prompted to report an adoption score $\hat{\alpha}_{pv}\in[0,1]$ for each parent $p\in\mathrm{Pa}(v)$. The score indicates the degree to which the current output relies on the information from parent $p$, rather than critiquing, revising, or rejecting it. Let $s_{pv}$ denote the reported adoption score. We set
\[
\hat{\alpha}_{pv}=s_{pv}, \qquad \hat{\beta}_{pv}=1-\hat{\alpha}_{pv}.
\]
In the online recurrence, $\hat{\alpha}_{pv}$ is used as the empirical edge weight in Eq.~\eqref{eq:local_recurrence}, serving as a proxy for the latent transmission parameter $\alpha_{pv}$.

This procedure introduces no additional training and can be applied to arbitrary MAS topologies. Since the elicited score is a self-reported proxy rather than a direct observation of the error-conditioned probability, its calibration may be imperfect. Nevertheless, the mathematical properties in Section~\ref{sec:properties}, such as boundedness, attenuation, and linear scalability, only require edge weights in $[0,1]$ and therefore also hold for the plug-in weights used during online inference. The prompt template is shown in Figure~\ref{fig:alpha_prompt_template}.

\section{Detailed Proofs and Derivations}
\label{app:properties}

This appendix provides the formal derivation of the uncertainty recurrence presented in Eq.~\eqref{eq:local_recurrence}, followed by detailed proofs of the key properties outlined in Section~\ref{sec:properties}.

\subsection{Derivation of the Uncertainty Recurrence}
\label{app:local_recurrence_derivation}

We derive the node-wise uncertainty recurrence in Eq.~\eqref{eq:local_recurrence}. Recall from the structural error composition in Eq.~\eqref{eq:error_logic} that the overall error event $E_v$ at node $v$ is defined as:
\[
E_v = I_v \vee \left( \bigvee_{p \in \mathrm{Pa}(v)} Z_{pv} \right).
\]
Applying De Morgan's laws, the complement event, representing a successful and error-free output at node $v$, occurs if and only if the agent avoids both local intrinsic reasoning failures and cascading external contamination:
\[
E_v = 0 \iff I_v = 0 \quad \text{and} \quad Z_{pv} = 0, \quad \forall p \in \mathrm{Pa}(v).
\]
Expressing this structural relationship in terms of joint probability yields:
\[
\Pr(E_v = 0) = \Pr\left( I_v = 0 \ \land \ \bigwedge_{p \in \mathrm{Pa}(v)} Z_{pv} = 0 \right).
\]
By invoking Assumption~\ref{asmp:intrinsic_independence} (Local-Propagation Independence), the local error event $I_v$ is conditionally independent of the incoming edge-level propagation events. This conditional independence allows us to factorize the joint probability as follows:
\[
\Pr(E_v = 0) = \Pr(I_v = 0) \cdot \Pr\left( \bigwedge_{p \in \mathrm{Pa}(v)} Z_{pv} = 0 \right).
\]
Next, by applying Assumption~\ref{asmp:edge_independence} (Edge-Wise Conditional Independence), the contamination events are mutually independent across distinct incoming edges. This tracking condition reduces the joint probability of the intersection to a product of marginal probabilities:
\[
\Pr\left( \bigwedge_{p \in \mathrm{Pa}(v)} Z_{pv} = 0 \right) = \prod_{p \in \mathrm{Pa}(v)} \Pr(Z_{pv} = 0).
\]
Combining these two factorizations yields the fully decoupled expression for the success probability:
\begin{equation}
\label{eq:success_prob_decom}
\Pr(E_v = 0) = \Pr(I_v = 0) \prod_{p \in \mathrm{Pa}(v)} \Pr(Z_{pv} = 0).
\end{equation}

By definition, the success probability of isolated local reasoning is $\Pr(I_v = 0) = 1 - u_v$. For each edge-level contamination event $Z_{pv}$, its marginal probability can be expanded by conditioning on the parent error state $E_p$:
\begin{align*}
\Pr(Z_{pv} = 1) &= \Pr(E_p = 1 \ \land \ M_{pv} = \mathrm{A}) \\
&= \Pr(M_{pv} = \mathrm{A} \mid E_p = 1) \Pr(E_p = 1) \\
&= \alpha_{pv} r_p,
\end{align*}
which directly implies $\Pr(Z_{pv} = 0) = 1 - \alpha_{pv} r_p$. Substituting these probability terms back into Eq.~\eqref{eq:success_prob_decom}, we obtain:
\[
\Pr(E_v = 0) = (1 - u_v) \prod_{p \in \mathrm{Pa}(v)} (1 - \alpha_{pv} r_p).
\]
Finally, utilizing the fundamental complement relation for total node uncertainty, $r_v = \Pr(E_v = 1) = 1 - \Pr(E_v = 0)$, we arrive at the exact closure:
\[
r_v = 1 - (1 - u_v) \prod_{p \in \mathrm{Pa}(v)} (1 - \alpha_{pv} r_p),
\]
which completes the proof of Eq.~\eqref{eq:local_recurrence}.

\paragraph{Boundary Condition for Source Nodes.} By mathematical convention, for any source node $v$ with no predecessors (i.e., $\mathrm{Pa}(v) = \emptyset$), the empty product evaluates to $1$. Under this condition, the recurrence naturally simplifies to $r_v = u_v$, preserving structural consistency across arbitrary graph entries.

\subsection{Proof of Proposition~\ref{prop:boundedness} (Boundedness)}
\label{app:proof_boundedness}

We verify that the local uncertainty recurrence relation preserves the algebraic constraints of a valid probability measure. Given the assumption that $u_v, r_p, \alpha_{pv} \in [0,1]$ for all $p \in \mathrm{Pa}(v)$, it follows that $0 \le \alpha_{pv}r_p \le 1$ for each incoming dependency edge $(p,v) \in \mathcal{E}$. This directly implies:
\[
0 \le 1 - \alpha_{pv}r_p \le 1.
\]
Since the unit interval $[0,1]$ is structurally closed under finite multiplication, the joint product over the parent configuration satisfies:
\[
0 \le \prod_{p \in \mathrm{Pa}(v)} (1 - \alpha_{pv}r_p) \le 1.
\]
Multiplying this product by the bounded local success probability $1 - u_v \in [0,1]$ preserves the interval containment:
\[
0 \le (1 - u_v) \prod_{p \in \mathrm{Pa}(v)} (1 - \alpha_{pv}r_p) \le 1.
\]
Finally, substituting this term back into the complementation formula of Eq.~\eqref{eq:local_recurrence} yields:
\[
r_v = 1 - (1 - u_v) \prod_{p \in \mathrm{Pa}(v)} (1 - \alpha_{pv}r_p) \implies 0 \le r_v \le 1.
\]
In the boundary case where $\mathrm{Pa}(v) = \emptyset$, the empty product evaluates to $1$, and the expression smoothly collapses to $r_v = 1 - (1 - u_v) = u_v$, ensuring perfect alignment with the boundary condition. This proves Proposition~\ref{prop:boundedness}.

\subsection{Proof of Proposition~\ref{prop:attenuation}: Uncertainty Attenuation}
\label{app:proof_attenuation}

Consider a non-source node $v$ such that $\mathrm{Pa}(v) \neq \emptyset$, and let $R_v = \max_{p \in \mathrm{Pa}(v)} r_p$ denote the maximum uncertainty among its parent nodes. To facilitate the algebraic analysis, we define the aggregate incoming contamination probability as:
\[
B_v := 1 - \prod_{p \in \mathrm{Pa}(v)} (1 - \alpha_{pv} r_p).
\]
By rearranging the terms of the core recurrence relation in Eq.~\eqref{eq:local_recurrence}, the node-wise uncertainty $r_v$ can be equivalently expressed as a linear combination of local uncertainty and external contamination:
\begin{equation}
\label{eq:rewritten_recurrence}
r_v = u_v + (1 - u_v) B_v.
\end{equation}

By hypothesis, we assume that the node possesses a lower local risk than its parents, $u_v < R_v$, and that its aggregate contamination is bounded by:
\begin{equation}
\label{eq:hypothesis_bound}
B_v < \frac{R_v - u_v}{1 - u_v}.
\end{equation}
Since $R_v \le 1$, the condition $u_v < R_v$ guarantees that $u_v < 1$, which ensures that the denominator $(1 - u_v)$ is strictly positive. Consequently, we can multiply both sides of the inequality in Eq.~\eqref{eq:hypothesis_bound} by $(1 - u_v)$ without reversing the inequality sign:
\[
(1 - u_v) B_v < R_v - u_v.
\]
Adding $u_v$ to both sides yields:
\[
u_v + (1 - u_v) B_v < R_v.
\]
Finally, substituting the rewritten recurrence relation from Eq.~\eqref{eq:rewritten_recurrence} into the left-hand side of the inequality directly produces:
\[
r_v < R_v.
\]
This demonstrates that a downstream node can achieve a lower uncertainty score than its predecessors, formally establishing that risk does not monotonically accumulate along the MAS execution trajectory. This completes the proof of Proposition~\ref{prop:attenuation}.

\begin{table}[t]
\centering
\caption{Qwen3-14B final-output uncertainty estimation across MAS topologies. \textsc{PropUQ} cells are shaded by relative change over each baseline: green/red indicates gains/drops, and darker colors indicate larger absolute changes.}
\label{tab:exp1-qwen3-14b-heatmap}
\fontsize{7.5pt}{8.7pt}\selectfont
\begingroup
\setlength{\tabcolsep}{3.5pt}
\renewcommand{\arraystretch}{1.5}
\begin{tabular}{l|l|c c|c c}
\hline
\multirow{2}{*}{\textbf{Dataset}} & \multirow{2}{*}{\textbf{Metric}} & \multicolumn{4}{c}{\textbf{Qwen3-14B}} \\
\cmidrule(lr){3-6}
 & & MSP & +\textsc{PropUQ} & Verb. & +\textsc{PropUQ} \\
\hline
\multicolumn{6}{>{\cellcolor[gray]{0.95}}c}{\textbf{\textit{Sequential MAS}}} \\
\hline
\multirow{2}{*}{\textbf{GSM8K}} & AUROC & 0.715 & \cellcolor[HTML]{FAE8E8}0.677 & 0.580 & \cellcolor[HTML]{C8E8C7}\textbf{0.651} \\
 & PRR & 0.403 & \cellcolor[HTML]{F8E0E0}0.374 & 0.159 & \cellcolor[HTML]{B3E0B2}\textbf{0.271} \\
\hline
\multirow{2}{*}{\textbf{MBPP+}} & AUROC & 0.690 & \cellcolor[HTML]{FAE9E9}0.655 & 0.544 & \cellcolor[HTML]{E9F6E8}\textbf{0.569} \\
 & PRR & 0.291 & \cellcolor[HTML]{D0ECCF}\textbf{0.321} & -0.005 & \cellcolor[HTML]{B3E0B2}\textbf{0.111} \\
\hline
\multirow{2}{*}{\textbf{MedQA}} & AUROC & 0.734 & \cellcolor[HTML]{C2E6C1}\textbf{0.835} & 0.588 & \cellcolor[HTML]{B3E0B2}\textbf{0.756} \\
 & PRR & 0.378 & \cellcolor[HTML]{B3E0B2}\textbf{0.601} & 0.178 & \cellcolor[HTML]{B3E0B2}\textbf{0.393} \\
\hline
\multicolumn{6}{>{\cellcolor[gray]{0.95}}c}{\textbf{\textit{Hierarchical MAS}}} \\
\hline
\multirow{2}{*}{\textbf{GSM8K}} & AUROC & 0.649 & \cellcolor[HTML]{CBEACA}\textbf{0.723} & 0.543 & \cellcolor[HTML]{F2FAF1}\textbf{0.557} \\
 & PRR & 0.237 & \cellcolor[HTML]{B3E0B2}\textbf{0.375} & 0.076 & \cellcolor[HTML]{B3E0B2}\textbf{0.105} \\
\hline
\multirow{2}{*}{\textbf{MBPP+}} & AUROC & 0.674 & \cellcolor[HTML]{D5EED5}\textbf{0.734} & 0.611 & \cellcolor[HTML]{E3F4E3}\textbf{0.647} \\
 & PRR & 0.278 & \cellcolor[HTML]{B3E0B2}\textbf{0.399} & 0.221 & \cellcolor[HTML]{BBE3BB}\textbf{0.255} \\
\hline
\multirow{2}{*}{\textbf{MedQA}} & AUROC & 0.662 & \cellcolor[HTML]{CCEACB}\textbf{0.737} & 0.681 & \cellcolor[HTML]{FDF5F5}0.664 \\
 & PRR & 0.239 & \cellcolor[HTML]{B3E0B2}\textbf{0.403} & 0.249 & \cellcolor[HTML]{FBEBEB}0.237 \\
\hline
\multicolumn{6}{>{\cellcolor[gray]{0.95}}c}{\textbf{\textit{Decentralized MAS}}} \\
\hline
\multirow{2}{*}{\textbf{GSM8K}} & AUROC & 0.608 & \cellcolor[HTML]{D8EFD7}\textbf{0.659} & 0.586 & \cellcolor[HTML]{EAF6EA}\textbf{0.612} \\
 & PRR & 0.200 & \cellcolor[HTML]{B3E0B2}\textbf{0.307} & 0.154 & \cellcolor[HTML]{B3E0B2}\textbf{0.181} \\
\hline
\multirow{2}{*}{\textbf{MBPP+}} & AUROC & 0.710 & \cellcolor[HTML]{F0F9F0}\textbf{0.732} & 0.575 & \cellcolor[HTML]{E4F4E3}\textbf{0.608} \\
 & PRR & 0.350 & \cellcolor[HTML]{D6EED6}\textbf{0.381} & 0.128 & \cellcolor[HTML]{B3E0B2}\textbf{0.174} \\
\hline
\multirow{2}{*}{\textbf{MedQA}} & AUROC & 0.753 & \cellcolor[HTML]{DAF0DA}\textbf{0.813} & 0.684 & \cellcolor[HTML]{FDFEFD}\textbf{0.686} \\
 & PRR & 0.401 & \cellcolor[HTML]{B3E0B2}\textbf{0.535} & 0.372 & \cellcolor[HTML]{F9E1E1}0.346 \\
\hline
\end{tabular}
\endgroup
\end{table}

\begin{table}[t]
\centering
\caption{Qwen3-14B intermediate-agent uncertainty estimation under decentralized MAS. \textsc{PropUQ} cells are shaded by relative change over each baseline: green/red indicates gains/drops, and darker colors indicate larger absolute changes.}
\label{tab:exp2-qwen3-14b}
\fontsize{7.2pt}{8.3pt}\selectfont
\begingroup
\setlength{\tabcolsep}{3.5pt}
\renewcommand{\arraystretch}{1.5}
\begin{tabular}{l|l|l|c c|c c}
\hline
\multirow{2}{*}{\textbf{Dataset}} & \multirow{2}{*}{\textbf{Agent}} & \multirow{2}{*}{\textbf{Metric}} & \multicolumn{4}{c}{\textbf{Qwen3-14B}} \\
\cmidrule(lr){4-7}
 & & & MSP & +\textsc{PropUQ} & Verb. & +\textsc{PropUQ} \\
\hline
\multirow[c]{4}{*}{\textbf{GSM8K}} & \multirow{2}{*}{Agent 2} & AUROC & 0.645 & \cellcolor[HTML]{F6FBF6}\textbf{0.656} & 0.564 & \cellcolor[HTML]{E6F5E6}\textbf{0.593} \\
 &  & PRR & 0.272 & \cellcolor[HTML]{C3E6C2}\textbf{0.309} & 0.100 & \cellcolor[HTML]{B3E0B2}\textbf{0.125} \\
\cline{2-7}
 & \multirow{2}{*}{Agent 3} & AUROC & 0.673 & \cellcolor[HTML]{F3FAF3}\textbf{0.689} & 0.516 & \cellcolor[HTML]{DBF0DA}\textbf{0.556} \\
 &  & PRR & 0.305 & \cellcolor[HTML]{BBE3BB}\textbf{0.352} & 0.037 & \cellcolor[HTML]{B3E0B2}\textbf{0.064} \\
\hline
\multirow[c]{4}{*}{\textbf{MBPP+}} & \multirow{2}{*}{Agent 2} & AUROC & 0.698 & \cellcolor[HTML]{E0F2DF}\textbf{0.742} & 0.584 & \cellcolor[HTML]{F7FCF7}\textbf{0.593} \\
 &  & PRR & 0.343 & \cellcolor[HTML]{D6EED5}\textbf{0.371} & 0.157 & \cellcolor[HTML]{D4EED4}\textbf{0.171} \\
\cline{2-7}
 & \multirow{2}{*}{Agent 3} & AUROC & 0.751 & \cellcolor[HTML]{F6FCF6}\textbf{0.764} & 0.621 & \cellcolor[HTML]{FFFEFE}0.620 \\
 &  & PRR & 0.434 & \cellcolor[HTML]{FBFDFB}\textbf{0.438} & 0.206 & \cellcolor[HTML]{D2EDD1}\textbf{0.226} \\
\hline
\multirow[c]{4}{*}{\textbf{MedQA}} & \multirow{2}{*}{Agent 2} & AUROC & 0.812 & \cellcolor[HTML]{F0F9F0}\textbf{0.837} & 0.674 & \cellcolor[HTML]{E3F4E3}\textbf{0.711} \\
 &  & PRR & 0.548 & \cellcolor[HTML]{E2F3E2}\textbf{0.581} & 0.370 & \cellcolor[HTML]{F8FCF8}\textbf{0.376} \\
\cline{2-7}
 & \multirow{2}{*}{Agent 3} & AUROC & 0.789 & \cellcolor[HTML]{E3F4E3}\textbf{0.834} & 0.719 & \cellcolor[HTML]{DFF2DE}\textbf{0.765} \\
 &  & PRR & 0.509 & \cellcolor[HTML]{CBEACA}\textbf{0.568} & 0.452 & \cellcolor[HTML]{FCEFEF}0.435 \\
\hline
\end{tabular}
\endgroup
\end{table}

\subsection{Proof of Proposition~\ref{prop:scalability}: Linear Scalability}
\label{app:proof_scalability}

By construction, the MAS execution trajectory is represented as a directed acyclic graph (DAG) $\mathcal{G}=(\mathcal{V},\mathcal{E})$, which inherently admits a topological sorting of its vertices, denoted by the sequence $v_1, v_2, \ldots, v_{|\mathcal{V}|}$. For any directed dependency edge $(p,v) \in \mathcal{E}$, the predecessor $p$ strictly precedes the successor $v$ within this sequence. This structural dependency guarantees that evaluating the recursive mapping in Eq.~\eqref{eq:local_recurrence} sequentially according to the topological order always ensures that all requisite parent uncertainties $\{r_p : p \in \mathrm{Pa}(v)\}$ are fully computed prior to visiting node $v$.

The local calculation for an individual node $v$ requires a single pass over its immediate parent set $\mathrm{Pa}(v)$:
\[
r_v = 1 - (1 - u_v) \prod_{p \in \mathrm{Pa}(v)} (1 - \alpha_{pv} r_p).
\]
The computational overhead for evaluating this node-wise recurrence scales linearly with its in-degree, yielding a local time complexity of $\mathcal{O}(|\mathrm{Pa}(v)|)$. Summing these individual operational costs across the entire vertex set $\mathcal{V}$ leverages a fundamental graph-theoretic identity:
\[
\sum_{v \in \mathcal{V}} \mathcal{O}(|\mathrm{Pa}(v)|) = \mathcal{O}\left( \sum_{v \in \mathcal{V}} |\mathrm{Pa}(v)| \right) = \mathcal{O}(|\mathcal{E}|).
\]
Given that finding or verifying the topological ordering via standard algorithms (e.g., Kahn's algorithm or Depth-First Search) incurs an execution cost of $\mathcal{O}(|\mathcal{V}| + |\mathcal{E}|)$, the cumulative computational time for tracking uncertainties across the entire MAS trajectory is tightly bounded by:
\[
\mathcal{O}(|\mathcal{V}| + |\mathcal{E}|).
\]

Furthermore, the auxiliary space complexity is $\mathcal{O}(|\mathcal{V}|)$ to store the scalar uncertainty values $\{r_v\}_{v \in \mathcal{V}}$, plus $\mathcal{O}(|\mathcal{V}| + |\mathcal{E}|)$ to maintain the underlying graph adjacency lists and edge-level interaction parameters. Consequently, both the time and space overheads scale linearly with the scale of MAS execution, completing the proof of Proposition~\ref{prop:scalability}.

\section{Supplementary Experimental Setup}
\subsection{Benchmark Descriptions}
\label{app:datasets}

\paragraph{GSM8K.}
GSM8K~\citep{gsm8k} contains 8.5K grade-school math word problems designed to evaluate multi-step numerical reasoning. Each problem requires translating a natural-language question into a sequence of arithmetic operations and producing a final numeric answer.

\paragraph{MedQA.}
MedQA~\citep{medqa} consists of medical licensing exam questions that test biomedical knowledge, clinical reasoning, and diagnostic decision-making. The benchmark requires models to integrate textual context with domain-specific medical knowledge under a multiple-choice setting.

\paragraph{MBPP-Plus.}
MBPP-Plus~\citep{mbppplus} extends the original MBPP benchmark with additional test cases and stricter execution-based evaluation. Each problem asks the model to generate a self-contained Python function that satisfies a functional specification.


\begin{table*}[t]
\centering
\caption{Mean uncertainty estimation under decentralized MAS. We report AUROC and PRR averaged over Agents 2--4; superscripts show relative changes for \textsc{PropUQ}, with green/red indicating gains/drops.}
\label{tab:exp2-mean}
\fontsize{7.5pt}{8.7pt}\selectfont
\begingroup
\setlength{\tabcolsep}{3.5pt}
\renewcommand{\arraystretch}{1.35}
\begin{tabular}{l|l|c c|c c|c c|c c|c c|c c}
\hline
\multirow{2}{*}{\textbf{Dataset}} & \multirow{2}{*}{\textbf{Metric}} & \multicolumn{4}{c}{\textbf{Qwen3-4B}} & \multicolumn{4}{|c}{\textbf{Qwen3-8B}} & \multicolumn{4}{|c}{\textbf{Gemma-3-12B}} \\
\cmidrule(lr){3-6} \cmidrule(lr){7-10} \cmidrule(lr){11-14}
 & & MSP & +\textsc{PropUQ} & Verb. & +\textsc{PropUQ} & MSP & +\textsc{PropUQ} & Verb. & +\textsc{PropUQ} & MSP & +\textsc{PropUQ} & Verb. & +\textsc{PropUQ} \\
\hline
\multirow{2}{*}{\textbf{GSM8K}} & AUROC & 0.664 & \textbf{0.697}\textsuperscript{\textbf{\textcolor[HTML]{006400}{+5.0\%}}} & 0.560 & \textbf{0.599}\textsuperscript{\textbf{\textcolor[HTML]{006400}{+7.0\%}}} & 0.735 & \textbf{0.766}\textsuperscript{\textbf{\textcolor[HTML]{006400}{+4.1\%}}} & 0.529 & \textbf{0.596}\textsuperscript{\textbf{\textcolor[HTML]{006400}{+12.8\%}}} & 0.571 & \textbf{0.575}\textsuperscript{\textbf{\textcolor[HTML]{006400}{+0.6\%}}} & 0.634 & \textbf{0.664}\textsuperscript{\textbf{\textcolor[HTML]{006400}{+4.7\%}}} \\
 & PRR & 0.351 & \textbf{0.423}\textsuperscript{\textbf{\textcolor[HTML]{006400}{+20.7\%}}} & 0.131 & \textbf{0.210}\textsuperscript{\textbf{\textcolor[HTML]{006400}{+60.7\%}}} & 0.453 & \textbf{0.524}\textsuperscript{\textbf{\textcolor[HTML]{006400}{+15.5\%}}} & 0.045 & \textbf{0.209}\textsuperscript{\textbf{\textcolor[HTML]{006400}{+361.0\%}}} & 0.146 & \textbf{0.166}\textsuperscript{\textbf{\textcolor[HTML]{006400}{+13.5\%}}} & 0.259 & \textbf{0.285}\textsuperscript{\textbf{\textcolor[HTML]{006400}{+10.1\%}}} \\
\hline
\multirow{2}{*}{\textbf{MBPP+}} & AUROC & 0.710 & \textbf{0.735}\textsuperscript{\textbf{\textcolor[HTML]{006400}{+3.5\%}}} & 0.595 & \textbf{0.608}\textsuperscript{\textbf{\textcolor[HTML]{006400}{+2.2\%}}} & 0.697 & \textbf{0.718}\textsuperscript{\textbf{\textcolor[HTML]{006400}{+3.0\%}}} & 0.522 & \textbf{0.577}\textsuperscript{\textbf{\textcolor[HTML]{006400}{+10.5\%}}} & 0.628 & \textbf{0.651}\textsuperscript{\textbf{\textcolor[HTML]{006400}{+3.7\%}}} & 0.666 & \textbf{0.705}\textsuperscript{\textbf{\textcolor[HTML]{006400}{+5.9\%}}} \\
 & PRR & 0.388 & \textbf{0.455}\textsuperscript{\textbf{\textcolor[HTML]{006400}{+17.1\%}}} & 0.173 & \textbf{0.210}\textsuperscript{\textbf{\textcolor[HTML]{006400}{+21.6\%}}} & 0.335 & \textbf{0.365}\textsuperscript{\textbf{\textcolor[HTML]{006400}{+8.8\%}}} & 0.049 & \textbf{0.135}\textsuperscript{\textbf{\textcolor[HTML]{006400}{+177.4\%}}} & 0.231 & \textbf{0.302}\textsuperscript{\textbf{\textcolor[HTML]{006400}{+30.8\%}}} & 0.347 & \textbf{0.394}\textsuperscript{\textbf{\textcolor[HTML]{006400}{+13.5\%}}} \\
\hline
\multirow{2}{*}{\textbf{MedQA}} & AUROC & 0.767 & \textbf{0.812}\textsuperscript{\textbf{\textcolor[HTML]{006400}{+5.9\%}}} & 0.679 & \textbf{0.728}\textsuperscript{\textbf{\textcolor[HTML]{006400}{+7.2\%}}} & 0.795 & \textbf{0.818}\textsuperscript{\textbf{\textcolor[HTML]{006400}{+2.9\%}}} & 0.693 & \textbf{0.831}\textsuperscript{\textbf{\textcolor[HTML]{006400}{+19.9\%}}} & 0.612 & \textbf{0.646}\textsuperscript{\textbf{\textcolor[HTML]{006400}{+5.5\%}}} & 0.584 & \textbf{0.615}\textsuperscript{\textbf{\textcolor[HTML]{006400}{+5.4\%}}} \\
 & PRR & 0.459 & \textbf{0.543}\textsuperscript{\textbf{\textcolor[HTML]{006400}{+18.5\%}}} & 0.334 & \textbf{0.348}\textsuperscript{\textbf{\textcolor[HTML]{006400}{+4.3\%}}} & 0.492 & \textbf{0.539}\textsuperscript{\textbf{\textcolor[HTML]{006400}{+9.6\%}}} & 0.386 & \textbf{0.566}\textsuperscript{\textbf{\textcolor[HTML]{006400}{+46.7\%}}} & 0.204 & \textbf{0.239}\textsuperscript{\textbf{\textcolor[HTML]{006400}{+17.1\%}}} & 0.209 & \textbf{0.229}\textsuperscript{\textbf{\textcolor[HTML]{006400}{+9.6\%}}} \\
\hline
\end{tabular}
\endgroup
\end{table*}

\begin{table*}[t]
\centering
\caption{Comparison with MATU and UProp on MedQA using Qwen3-8B. Time is in seconds per example, and UProp applies only to sequential MAS. Bold marks the best uncertainty metric within each topology.}
\label{tab:concurrent_sampling_baselines}
\vspace{-0.2cm}
\small
\begingroup
\setlength{\tabcolsep}{5.0pt}
\renewcommand{\arraystretch}{1.3}
\begin{tabular}{l|c c|c c|c c}
\hline
\multirow{2}{*}{\textbf{Metric}} & \multicolumn{2}{c|}{\textbf{Sampling-Based}} & \multicolumn{2}{c|}{\textbf{MSP}} & \multicolumn{2}{c}{\textbf{Verb.}} \\
\cmidrule(lr){2-3} \cmidrule(lr){4-5} \cmidrule(lr){6-7}
 & MATU & UProp & w/o & +\textsc{PropUQ} & w/o & +\textsc{PropUQ} \\
\hline
\multicolumn{7}{>{\cellcolor[gray]{0.95}}c}{\textbf{\textit{Sequential MAS}}} \\
\hline
AUROC & 0.718 & 0.650 & 0.754 & \textbf{0.842} & 0.600 & 0.791 \\
PRR & 0.373 & 0.330 & 0.496 & \textbf{0.624} & 0.217 & 0.482 \\
Time/sample (s) & 41.410 & 11.310 & 1.129 & 1.328 & 4.059 & 4.258 \\
\hline
\multicolumn{7}{>{\cellcolor[gray]{0.95}}c}{\textbf{\textit{Hierarchical MAS}}} \\
\hline
AUROC & 0.626 & -- & 0.658 & 0.711 & 0.604 & \textbf{0.722} \\
PRR & 0.263 & -- & 0.277 & 0.349 & 0.223 & \textbf{0.362} \\
Time/sample (s) & 56.040 & -- & 1.536 & 1.735 & 5.522 & 5.721 \\
\hline
\end{tabular}
\endgroup
\end{table*}

\begin{figure*}[htbp]
    \centering
    \includegraphics[width=\linewidth]{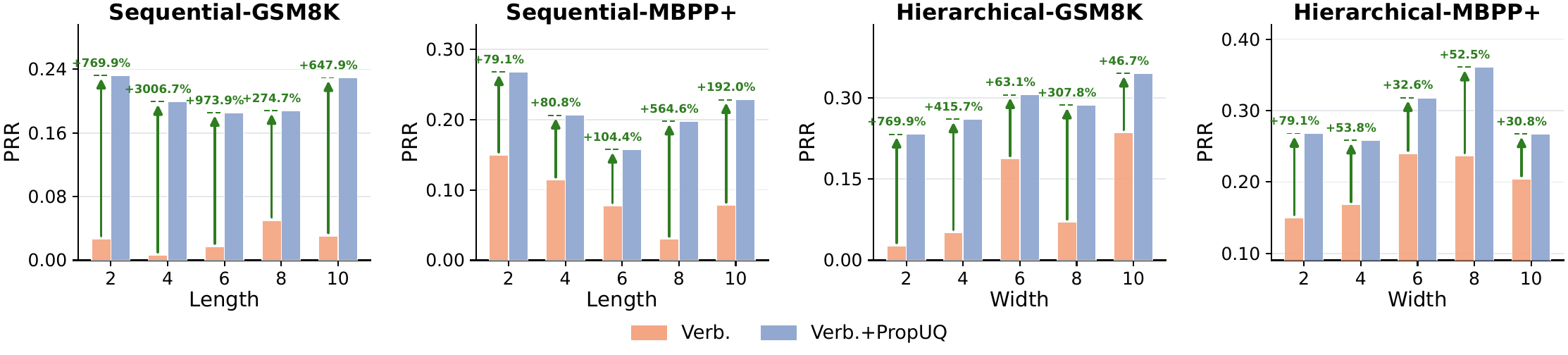} 
    \caption{
    Structural scale generalization results measured by PRR. We evaluate the robustness of uncertainty propagation across varying sequential chain lengths and hierarchical expert-layer widths.
    }
    \label{fig:generalization_prr}
\end{figure*}

\subsection{MAS Structure Details}
\label{app:mas_structures}

We provide additional details for the three MAS interaction structures used in our experiments.

\paragraph{Sequential MAS.}
In the \textbf{sequential MAS}, we adopt a chain-of-agents design~\citep{chain1,chain2} with four role-specialized agents: planner, critic, refiner, and solver. The planner first receives the input question, and each downstream agent processes the reasoning output of its immediate predecessor.

\paragraph{Hierarchical MAS.}
In the \textbf{hierarchical MAS}, we follow a domain-specialized design~\citep{hierarchical1,hierarchical2}, where code, math, and science agents independently reason from different domain perspectives, and a summarizer agent aggregates their intermediate responses to produce the final answer.

\paragraph{Decentralized MAS.}
In the \textbf{decentralized MAS}, agents have no predefined roles and communicate through a peer-to-peer mesh~\citep{mesh1,mesh2}. Each agent can condition on the input question and messages from the other agents, allowing us to evaluate uncertainty propagation under dense, role-free interactions.

\subsection{Evaluation Metrics}
\label{app:metrics}

Let $r_v \in [0,1]$ be the uncertainty score and $e_v \in \{0,1\}$ be the ground-truth error indicator ($e_v=1$ for incorrect outputs) for node $v \in \mathcal{V}_{\mathrm{eval}}$.

\paragraph{AUROC.} 
AUROC quantifies the probability that an incorrect node is ranked higher than a correct one. Let $\mathcal{P} = \{v: e_v=1\}$ and $\mathcal{N} = \{v: e_v=0\}$:
\begin{equation*}
\begin{aligned}
\mathrm{AUROC} = {} & \frac{1}{|\mathcal{P}||\mathcal{N}|} \sum_{v \in \mathcal{P}} \sum_{w \in \mathcal{N}} \Big[ \mathbf{1}(r_v > r_w) \\
                    & + \frac{1}{2}\mathbf{1}(r_v = r_w) \Big].
\end{aligned}
\end{equation*}

\paragraph{Prediction Rejection Ratio (PRR).} 
PRR evaluates uncertainty-informed \textit{selective prediction}. By sorting nodes in descending order of $r_v$, we compute the Area Under the Accuracy-Rejection Curve (AURC), denoted as $\mathrm{AUC}_{\mathrm{uq}}$. To ensure a metric independent of the base accuracy, PRR normalizes this area against random and oracle baselines:
\[
\mathrm{PRR} = \frac{\mathrm{AUC}_{\mathrm{uq}} - \mathrm{AUC}_{\mathrm{rand}}}{\mathrm{AUC}_{\mathrm{oracle}} - \mathrm{AUC}_{\mathrm{rand}}}.
\]
Higher AUROC and PRR scores signify superior discriminative power and practical utility for error filtering.

\section{Additional Experiment Results}
\label{app:additional_experiments}

\subsection{Final-output Results on Qwen3-14B.}
Table~\ref{tab:exp1-qwen3-14b-heatmap} reports additional final-output UQ results using \texttt{Qwen3-14B}. The experiments follow the same datasets, MAS topologies, metrics, and local UQ baselines used in the main evaluation in Section~\ref{sec:final_output_uq}. The results show that \textsc{PropUQ-MAS} remains effective at a larger Qwen3 scale, providing further evidence that the proposed propagation rule is not limited to smaller base models.

\subsection{Additional Intermediate-Step Results}
Table~\ref{tab:exp2-qwen3-14b} further reports the intermediate-step UQ results on \texttt{Qwen3-14B} under the decentralized MAS setting. Following the main intermediate-step evaluation in Section~\ref{sec:node_level_uq}, we report results for intermediate agents whose outputs can be directly judged against the ground-truth answer. \textsc{PropUQ-MAS} improves AUROC and PRR in most settings for both MSP and verbalized uncertainty baselines, indicating that the propagation rule also provides effective intermediate-step reliability signals at the larger Qwen3 scale.

Appendix Table~\ref{tab:exp2-mean} further summarizes the averaged intermediate-step results over Agents~2--4. The improvements are consistent across datasets, model families, and local UQ baselines: \textsc{PropUQ-MAS} improves AUROC in all reported settings and improves PRR in nearly all settings, with especially large PRR gains for verbalized uncertainty. These averaged results confirm that propagation-aware UQ provides stable intermediate-step reliability gains, rather than improvements limited to a single agent or dataset.

\subsection{Comparison with Sampling-Based Baselines}
\label{app:concurrent_sampling_baselines}

We compare \textsc{PropUQ-MAS} with two concurrent sampling-based UQ baselines, MATU~\citep{chen2026every} and UProp~\citep{duan2025uprop}, on MedQA using Qwen3-8B under sequential and hierarchical MAS. UProp is designed for single-agent trajectories and can therefore be adapted only to the sequential setting. At the time of our experiments, official implementations were unavailable; we therefore implemented both baselines following their original papers.

Table~\ref{tab:concurrent_sampling_baselines} shows that \textsc{PropUQ-MAS} achieves the strongest uncertainty estimation in both topologies: it outperforms MATU and UProp on sequential MAS, and both propagated variants outperform MATU on hierarchical MAS. Moreover, \textsc{PropUQ-MAS} is substantially more efficient than sampling-based baselines, as it introduces only a lightweight propagation layer during standard MAS execution. These results confirm that \textsc{PropUQ-MAS} effectively provides robust, online reliability signals with minimal computational overhead.

\subsection{PRR results for scale generalization}
Figure~\ref{fig:generalization_prr} reports the PRR results for the scale generalization analysis in Section~\ref{sec:generalization}. We vary the chain length in sequential MAS and the expert-layer width in hierarchical MAS. Across GSM8K and MBPP-Plus, \textsc{PropUQ-MAS} consistently improves over the verbalized uncertainty baseline, showing that the proposed propagation rule remains useful when the MAS trajectory becomes longer or when an aggregation node receives more parent inputs. These results complement the AUROC results in the main text and further support the scalability of \textsc{PropUQ-MAS} under both deeper and wider MAS structures.

\begin{figure}[t]
\centering
\begin{tcolorbox}[
    width=\linewidth,
    colback=gray!4!white,
    colframe=gray!70!black,
    colbacktitle=gray!70!black,
    coltitle=white,
    title=Additional Case Study from GSM8K under Sequential MAS,
    fonttitle=\bfseries\footnotesize,
    left=5pt, right=5pt, top=4pt, bottom=4pt,
    boxsep=2pt,
    arc=2pt
]
\footnotesize
\setlength{\parskip}{2pt}
\setlength{\parindent}{0pt}

\textbf{Question.}
There are 36 penguins sunbathing. One-third swim away, and another one-third go inside a cave. How many penguins are still sunbathing?

\textbf{Gold Answer:} \textcolor{ForestGreen}{12}
\hfill
\textbf{Final Prediction:} \textcolor{BrickRed}{16 (incorrect)}

\noindent\hdashrule{\linewidth}{0.4pt}{2pt 2pt}

\vspace{2pt}
\textbf{Planner.}
\textcolor{ForestGreen}{
\textit{Plan:} start with 36 penguins; compute one-third of 36 as 12 swimmers; compute another one-third of 36 as 12 cave-going penguins; subtract both groups from 36, giving \(36-12-12=12\).
}

\textit{Local uncertainty:} 0.15\\
\textit{Propagation-aware uncertainty:} 0.15

\vspace{1pt}
\begin{center}
\(\downarrow\) \quad Planner to Critic: acceptance = 0.60
\end{center}

\textbf{Critic.}
\textcolor{BrickRed}{
\textit{I think the plan may be incomplete:} ``another one-third'' could refer to one-third of the remaining 24 penguins rather than one-third of the original 36. Under this interpretation, \(24/3=8\), leading to \(24-8=16\).
}

\textit{Local uncertainty:} 0.75\\
\textit{Propagation-aware uncertainty:} 0.773

\vspace{1pt}
\begin{center}
\(\downarrow\) \quad Critic to Refiner: acceptance = 0.95
\end{center}

\textbf{Refiner.}
\textcolor{BrickRed}{
\textit{Refined plan:} start with 36; subtract 12 swimmers to get 24 remaining; then take one-third of the 24 as 8 cave-going penguins; therefore \(24-8=16\).
}

\textit{Local uncertainty:} 0.20\\
\textit{Propagation-aware uncertainty:} 0.787

\vspace{1pt}
\begin{center}
\(\downarrow\) \quad Refiner to Final Solver: acceptance = 0.90
\end{center}

\textbf{Final Solver.}
\textcolor{BrickRed}{
\textit{I follow the Refiner's plan:} after 12 penguins swim away, 24 remain; taking one-third of 24 gives 8, so the final answer is \(24-8=16\). The steps seem internally consistent, so my local uncertainty is low.
}

\textit{Local uncertainty:} 0.10\\
\textit{Propagation-aware uncertainty:} 0.738

\end{tcolorbox}
\caption{
Additional sequential MAS case study from GSM8K. The planner initially gives a correct plan, but the critic introduces an incorrect interpretation that is strongly adopted by later agents. Although the final solver has low local uncertainty, \textsc{PropUQ-MAS} assigns high propagation-aware uncertainty by tracking the accepted upstream risk.
}
\label{fig:case_study_sequential}
\end{figure}

\subsection{Sequential MAS Case Study}
\label{app:sequential_case_study}

Figure~\ref{fig:case_study_sequential} presents a GSM8K case under the sequential MAS topology. While the hierarchical case in Section~\ref{sec:case_study} illustrates selective aggregation over multiple upstream branches, this example focuses on a different failure mode: an error introduced at one intermediate step can be successively adopted and carried forward along a reasoning chain.

In this example, the planner first produces a correct plan. The critic then introduces an alternative but incorrect interpretation of the problem, which is strongly adopted by the refiner. The final solver follows the refined plan and outputs an incorrect answer despite having low local uncertainty. This shows that a downstream agent may appear reliable from its own local perspective while still inheriting an earlier reasoning error.

\textsc{PropUQ-MAS} identifies this failure by propagating uncertainty along the accepted dependencies in the sequential chain. As the incorrect interpretation is repeatedly accepted by later agents, the final output receives high propagation-aware uncertainty. This complementary case demonstrates that \textsc{PropUQ-MAS} can capture not only selective risk aggregation in hierarchical MAS, but also chain-wise risk transmission in sequential MAS.

\section{Prompt Templates}
\label{app:prompts}

This appendix provides the prompt templates used in our experiments.

\subsection{MAS Agent Response Prompts}
\label{app:mas_agent_prompt}

Figure~\ref{fig:sequential_prompt_template} and Figure~\ref{fig:hierarchical_prompt_template} show the role-specific prompts used for sequential and hierarchical MAS execution.

\begin{figure*}[htbp]
\centering
\begin{tcolorbox}[
    colback=gray!5!white,
    colframe=gray!75!black,
    title=Sequential MAS Role Prompts,
    fonttitle=\bfseries,
    left=5pt, right=5pt, top=5pt, bottom=5pt
]
\small
\textbf{Shared input:} \\
\texttt{\#\# Input Question: \{QUESTION\}} \\[3pt]

\textbf{Planner Agent:} \\
\texttt{You are a Planner Agent. Given the input question, design a concise step-by-step plan for solving it. Do not produce the final answer.} \\
\texttt{\#\# Output Format:} \\
\texttt{Planner Agent's Output: [Your plan here]} \\[5pt]

\textbf{Critic Agent:} \\
\texttt{You are a Critic Agent. You are given the input question and the Planner Agent's plan. Evaluate whether the plan is correct and complete, and provide helpful feedback.} \\
\texttt{\#\# Plan from Planner Agent: \{PLANNER\_OUTPUT\}} \\
\texttt{\#\# Output Format:} \\
\texttt{Critic Agent's Output:} \\
\texttt{Original Plan: [Copy the provided plan here]} \\
\texttt{Feedback: [Your feedback here]} \\[5pt]

\textbf{Refiner Agent:} \\
\texttt{You are a Refiner Agent. You are given the input question, the Planner Agent's plan, and the Critic Agent's feedback. Produce an improved step-by-step plan.} \\
\texttt{\#\# Original Plan and Critic Feedback: \{PLANNER\_AND\_CRITIC\_OUTPUT\}} \\
\texttt{\#\# Output Format:} \\
\texttt{Refiner Agent's Output: [Your refined plan here]} \\[5pt]

\textbf{Solver Agent:} \\
\texttt{You are the final Solver Agent in a sequential MAS: Planner $\rightarrow$ Critic $\rightarrow$ Refiner $\rightarrow$ Solver. You are given the input question and the Refiner Agent's plan. The plan may contain irrelevant or incorrect content; ignore it if it is not useful.} \\
\texttt{\#\# Refined Plan: \{REFINER\_OUTPUT\}} \\
\texttt{Reason step by step and output the final answer inside \textbackslash boxed\{YOUR\_FINAL\_ANSWER\}.}
\end{tcolorbox}
\caption{Role-specific prompt template used for the sequential MAS.}
\label{fig:sequential_prompt_template}
\end{figure*}

\begin{figure*}[htbp]
\centering
\begin{tcolorbox}[
    colback=gray!5!white,
    colframe=gray!75!black,
    title=Hierarchical MAS Role Prompts,
    fonttitle=\bfseries,
    left=5pt, right=5pt, top=5pt, bottom=5pt
]
\small
\textbf{Shared input:} \\
\texttt{\#\# Input Question: \{QUESTION\}} \\[3pt]

\textbf{Math Agent:} \\
\texttt{You are a math agent. Solve the input question from a mathematical reasoning perspective. Output the final answer inside \textbackslash boxed\{YOUR\_FINAL\_ANSWER\}.} \\[5pt]

\textbf{Science Agent:} \\
\texttt{You are a science agent. Solve the input question from a scientific reasoning perspective. Output the final answer inside \textbackslash boxed\{YOUR\_FINAL\_ANSWER\}.} \\[5pt]

\textbf{Code Agent:} \\
\texttt{You are a code agent. Solve the input question using programming or algorithmic reasoning when useful. Output the final answer inside \textbackslash boxed\{YOUR\_FINAL\_ANSWER\}.} \\[5pt]

\textbf{Task Summarizer:} \\
\texttt{You are a task summarizer. You are given the input question and the responses from the Math, Science, and Code agents. Synthesize their outputs and produce the final answer.} \\
\texttt{\#\# Previous Agent Outputs: \{MATH\_SCIENCE\_CODE\_OUTPUTS\}} \\
\texttt{Output the final answer inside \textbackslash boxed\{YOUR\_FINAL\_ANSWER\}.}
\end{tcolorbox}
\caption{Role-specific prompt template used for the hierarchical MAS.}
\label{fig:hierarchical_prompt_template}
\end{figure*}

\begin{figure*}[htbp]
\centering
\begin{tcolorbox}[
    colback=gray!5!white,
    colframe=gray!75!black,
    title=Self-Reporting Prompt for $\hat \alpha_{pv}$ Elicitation,
    fonttitle=\bfseries,
    left=5pt, right=5pt, top=5pt, bottom=5pt
]
\small
\texttt{For each incoming agent, report message\_adoption\_weight as a number in [0,1].} \\

\begin{itemize}[leftmargin=15pt, nosep]
    \item \small \texttt{A higher value means your response largely accepts or reuses that agent's message.}
    \item \small \texttt{A lower value means your response critiques, revises, rejects, or only weakly relies on that message.}\\
\end{itemize}

\texttt{\#\# Output one line for each directly connected incoming agent:} \\
\texttt{<message\_adoption agent="\{INCOMING\_AGENT\_NAME\}" score="FLOAT\_0\_TO\_1"/>}
\end{tcolorbox}
\caption{Prompt template used to elicit the self-reported acceptance score $\hat{\alpha}_{pv}$ after an agent response is generated.}
\label{fig:alpha_prompt_template}
\end{figure*}

\subsection{Acceptance Probability Elicitation Prompt}
\label{app:alpha_prompt}

Figure~\ref{fig:alpha_prompt_template} shows the prompt template used to elicit the self-reported acceptance score $\hat{\alpha}_{pv}$. This score serves as the plug-in estimate of the edge-level acceptance probability ${\alpha}_{pv}$ in online inference.

\subsection{Prompt for Scale Generalization Analysis}
\label{app:scaling_prompt}

In the scale generalization analysis, we vary the number of agents while keeping their behavior comparable across scales. Each agent follows the same critic-solver instruction: it verifies incoming answers, provides feedback, and then produces its own answer. This controlled design allows us to change the MAS depth or width without introducing additional role-specific effects. Figure~\ref{fig:scaling_prompt_template} shows the prompt used in this experiment.

\begin{figure*}[htbp]
\centering
\begin{tcolorbox}[
    colback=gray!5!white,
    colframe=gray!75!black,
    title=Critic-Solver Prompt for Scale Generalization,
    fonttitle=\bfseries,
    left=5pt, right=5pt, top=5pt, bottom=5pt
]
\small

\texttt{Given the input question and answers from directly connected previous agents, independently verify each incoming answer, decide whether you agree, and identify what should be improved. If you disagree, provide an alternative solution. Then provide feedback and your own answer.} \\

\texttt{\{ANSWER\_FORMAT\_INSTRUCTION\}} \\

\texttt{\#\# Output Format} \\
\texttt{\#\# Feedback} \\
\texttt{- Agent X: [State whether you agree, what is correct, and what should be improved]} 

\end{tcolorbox}
\caption{Prompt template used in the scale generalization analysis, where each agent acts as both critic and solver.}
\label{fig:scaling_prompt_template}
\end{figure*}

\subsection{LLM-as-a-Judge Prompt for Intermediate-Step Evaluation}
\label{app:node_judge_prompt}

For the intermediate-step evaluation in the sequential MAS, GPT-5.5 judges whether each Critic or Refiner output contains a substantive medical or reasoning error. Figure~\ref{fig:node_judge_prompt} shows the prompt template.

\begin{figure*}[htbp]
\centering
\begin{tcolorbox}[
    colback=gray!5!white,
    colframe=gray!75!black,
    title=LLM-as-a-Judge Prompt for Intermediate-Step Evaluation,
    fonttitle=\bfseries,
    left=5pt, right=5pt, top=5pt, bottom=5pt
]
\footnotesize
\textbf{System Message} \\
\texttt{You are a strict but fair expert judge for medical question answering.} \\
\texttt{Return exactly one valid JSON object and no additional text. /no\_think} \\[3pt]

\textbf{User Message} \\
\texttt{Evaluate the specified intermediate MAS node using the task, reference answer,} \\
\texttt{and preceding node outputs below. Assess the target node itself rather than} \\
\texttt{inferring its quality solely from the final MAS prediction.} \\[3pt]

\texttt{\#\#\# Task} \\
\texttt{Question: \{TASK\_QUESTION\}} \\
\texttt{Gold answer option: \{GOLD\_ANSWER\_OPTION\}} \\
\texttt{Final MAS prediction: \{FINAL\_MAS\_PREDICTION\}} \\[2pt]
\texttt{\#\#\# Previous Intermediate Context} \\
\texttt{\{PREVIOUS\_INTERMEDIATE\_CONTEXT\}} \\[2pt]
\texttt{\#\#\# Target Node} \\
\texttt{Role: \{NODE\_ROLE\}} \\
\texttt{Output: \{NODE\_OUTPUT\}} \\[2pt]
\texttt{\#\#\# Role-Specific Criterion} \\
\texttt{\{ROLE\_SPECIFIC\_JUDGING\_INSTRUCTION\}} \\[3pt]

\texttt{\#\#\# Label Definitions} \\
\begin{itemize}[leftmargin=14pt, nosep]
    \item \texttt{"helpful"}: medically valid and useful, without a substantive error.
    \item \texttt{"mixed"}: useful but incomplete, ambiguous, or affected by a limited issue.
    \item \texttt{"harmful"}: contains a substantive medical or reasoning error likely to mislead downstream agents.
\end{itemize}
\texttt{Set "node\_error" to true if and only if the label is "harmful"; otherwise, set it to false.} \\
\texttt{Return exactly one JSON object with only "label", "node\_error", and "rationale".} \\
\texttt{Do not include Markdown fences or any other text. /no\_think} \\[3pt]

\textbf{Role-Specific Criteria} \\
\texttt{Critic: Determine whether the Critic gives a medically valid critique of the preceding} \\
\texttt{plan. It should identify genuine issues or appropriately confirm correct reasoning} \\
\texttt{without introducing a substantive error.} \\
\texttt{Refiner: Determine whether the Refiner produces a medically valid revised plan. It} \\
\texttt{should preserve or improve the preceding reasoning without introducing a substantive error.} \\[3pt]

\textbf{Placeholder Definitions} \\
\begin{itemize}[leftmargin=14pt, nosep]
    \item \texttt{\{TASK\_QUESTION\}}: complete multiple-choice medical question, including answer options.
    \item \texttt{\{GOLD\_ANSWER\_OPTION\}}: reference answer option; \texttt{\{FINAL\_MAS\_PREDICTION\}}: final MAS answer.
    \item \texttt{\{PREVIOUS\_INTERMEDIATE\_CONTEXT\}}: preceding outputs in execution order; use \texttt{[none]} when unavailable.
    \item \texttt{\{NODE\_ROLE\}}: target role (Critic or Refiner); \texttt{\{NODE\_OUTPUT\}}: target output.
    \item \texttt{\{ROLE\_SPECIFIC\_JUDGING\_INSTRUCTION\}}: criterion for the target role.
\end{itemize}
\end{tcolorbox}
\caption{LLM-as-a-judge prompt for labeling Critic and Refiner outputs in the sequential MedQA evaluation.}
\label{fig:node_judge_prompt}
\end{figure*}

\end{document}